\documentclass[11pt,a4paper]{JHEP3}
\usepackage{amsmath, amssymb}
\usepackage{verbatim}
\usepackage{latexsym}

\def\be{\begin{eqnarray}}
 \def\ee{\end{eqnarray}}
 \def\0{\nonumber}
\def\d{\partial}
\usepackage{euscript}

\usepackage{amsfonts}
\usepackage{verbatim}

\newcommand\E{{\cal E}}

\def\g{{\rm g}}
\def\e{\epsilon}

\def\del{\partial}

\def\del{\partial}

\def\cos{{\rm cos}}
\def\sin{{\rm sin}}

\def\ket#1{|#1 \rangle}
\def\0{\nonumber}

\preprint{SISSA/88/2010/EP\\\tt hep-th/1105.5926}

\title{The energy of the analytic lump solution in SFT}

\author{ L.Bonora\\
International School for Advanced Studies (SISSA)\\
Via Bonomea 265, 34136 Trieste, Italy, and INFN, Sezione di
Trieste, Italy;\\
Yukawa Institute for Theoretical Physics, Kyoto University,\\
Kyoto 606-8502,Japan\\
E-mail:   \email{bonora@sissa.it},}

\author{ S.Giaccari\\
International School for Advanced Studies (SISSA)\\
Via Bonomea 265, 34136 Trieste, Italy, and INFN, Sezione di
Trieste, Italy;\\
E-mail:   \email{giaccari@sissa.it},}

\author{D.D.Tolla\\
Department of Physics and University College,
Sungkyunkwan University,
Suwon 440-746, South Korea\\
E-mail:  \email{ddtolla@skku.edu}}

\abstract{In a previous paper a method was proposed to find exact
analytic solutions of open string field theory describing lower
dimensional lumps, by incorporating in string field theory an exact
renormalization group flow generated by a relevant operator in a
worldsheet CFT. In this paper we compute the energy of one such solution, which is expected to represent a D24 brane. We show, both numerically and analytically, that its value 
corresponds to the theoretically expected one.}

\keywords{String Field Theory, Tachyon Condensation}

\begin{document}

\section{Introduction}

 In a recent paper, \cite{BMT}, which will be referred to as I, following an earlier suggestion of \cite{Ellwood}, a general method was described to
 obtain new exact analytic solutions in Witten's cubic open string field theory
(OSFT)~\cite{Witten:1985cc}, and in particular solutions that
describe inhomogeneous tachyon condensation. Let us recall that
these solutions fill the gap left in the verification of the
expectation that an OSFT defined on a particular boundary conformal
field theory (BCFT) has classical solutions describing other
boundary conformal field theories~\cite{Sen:1999mh,Sen:1999xm}. The
previous construction of analytic solutions describing the tachyon
vacuum~\cite{Schnabl05, Okawa1, ErlerSchnabl, RZ06, ORZ, Fuchs0,
Erler:2006hw, Erler:2006ww, Erler:2007xt, Arroyo:2010fq,
Zeze:2010jv, Zeze:2010sr, Arroyo:2010sy,Murata} and of those describing a
general marginal boundary deformations of the initial
BCFT~\cite{KORZ, Schnabl:2007az, Kluson:2003xu, Kiermaier:2007vu, Fuchs3,
Lee:2007ns, Kwon:2008ap, Okawa2, Okawa3, Kiermaier:2007ki,
Erler:2007rh}, had added ground for this expectation\footnote {See
\cite{Fuchs4, Schnabl:2010tb} for reviews.}. In all these
developments there was a missing element: the solutions describing
inhomogeneous and relevant boundary deformations of the initial BCFT
were not known, though their existence was
predicted~\cite{Sen:1999mh,Sen:1999xm,lumps}. In I such solutions
were put forward.

The method of I consists in translating an exact
renormalization group (RG) flow, generated in a two--dimensional world--sheet theory by a relevant operator,
to the language of OSFT. The so-constructed solution is a deformation of the Erler-Schnabl solution,
\cite{ErlerSchnabl}, the latter being a solution that describes homogeneous tachyon condensation for the D25 brane. It was shown in I that, if the operator has suitable properties, such solution will describe tachyon condensation in specific space directions, thus representing the condensation of a lower dimensional brane. In this paper we will analyze a particular
solution, generated by an exact RG flow analyzed first by Witten,
\cite{Witten}. In I we concluded that, on the basis of the analysis
carried out in the framework of 2D CFT in \cite{Kutasov}, this
solution should describe a D24 brane, with the correct ratio of
tension with respect to the starting D25 brane. Of course an
important piece of evidence for this interpretation is a precise
determination of the energy of the solution. This is our aim in this
paper.

As it happens, the expression of the energy for our solution in the
SFT language is very  complicated and does not allow for a
straightforward analytic evaluation. Nevertheless in this paper we
will be able to determine it exactly via an indirect method. As one
may suspect, the entire procedure is rather roundabout, so we would
like to spend some time explaining it. We start form a solution
$\psi_u$ of SFT {\it on the perturbative vacuum} and our first aim
is to show that its energy is finite. More precisely, the energy of
such solution has an UV ($u=0$) singularity, which originates from the
infinite volume factor due to our normalization, and corresponds to
the tachyon vacuum energy.  Once we have subtracted it, the energy
becomes finite and well defined. Specifically, the lump energy being
determined by an integration over a real variable $U$ from 0 to
$\infty$, we show that asymptotically the integrand behaves like
$1/U^2$, so that it is integrable. This is a piece of information
that will be used throughout.  Not only. Having pushed the analytic
calculation as far as possible, we will continue with the numerical
evaluation of the energy, obtaining finally a rather precise
numerical result, which differs by about 1/3 from the expected
theoretical value of a D24 brane tension. This teaches us that we
should not expect to find the right lump energy in a solution based
on the perturbative vacuum, whose energy functional depends on the UV subtraction.
The true energy must be independent of the latter.

Although the previous numerical calculation is not satisfactory it
will turn out instrumental in the sequel. But, in order to be able
to access an analytic evaluation of the energy,  we have to take a
detour. To this end we introduce an $\e$ regulator in the Schwinger
representation we use to represent the solution. The so--obtained
regularized solution is called $\psi_\e$, the original solution
being identified with the $\e=0$ one.  We will show that $\psi_\e$
is a tachyon condensation vacuum solution, whose energy, after
subtracting the UV contribution, is expected to be exactly 0. We
will show that this is indeed the case to a great accuracy. This
will allow us to evaluate the energy of the original solution to a
great accuracy too, confirming in fact the previous numerical
calculation.

At this point everything is ready for the exact calculation of the
lump energy.  The lump string field is not the initial $\psi_u$, but
$\psi_u-\psi_\e$, which is a solution to the SFT equation of motion
{\it on the tachyon vacuum}. The analytic determination of its
energy is almost elementary, is UV subtraction--independent and gives the expected theoretical
result.

To start with let us briefly summarize the construction of I.

\subsection{Review of the results from I}

In I, to start with, the well-known $K,B,c$ algebra
defined by 
\be 
K=\frac\pi2K_1^L\ket I, \quad \quad
B=\frac\pi2B_1^L\ket I,\quad\quad c= c\left(\frac12\right)\ket I,
\label{KBc} 
\ee in the sliver frame (obtained by mapping the UHP to
an infinite cylinder $C_2$ of circumference 2, by the sliver map
$\tilde z=\frac2\pi\arctan z$), was enlarged by adding a (relevant) matter operator
\be \phi=\phi\left(\frac12\right)\ket I\label{phi} \ee with the
properties \be \,[c,\phi]=0,\quad\quad \,[B,\phi]=0,\quad\quad
\,[K,\phi]= \del\phi,\label{proper} 
\ee 
such that $Q$ has the following action: 
\be 
Q\phi=c\del\phi+\del c\delta\phi.
\label{actionQ} 
\ee 
It can be easily proven that 
\be 
\psi_{\phi}=c\phi-\frac1{K+\phi}(\phi-\delta\phi) Bc\del c\label{psiphi} 
\ee
does indeed satisfy the OSFT equation of motion 
\be Q\psi_{\phi}+\psi_{\phi}\psi_{\phi}=0\label{eom}. 
\ee 
It is clear
that (\ref{psiphi}) is a deformation of the Erler--Schnabl solution,
see \cite{ErlerSchnabl}, which can be recovered for $\phi=1$.

Much like in the Erler-Schnabl (ES) case, we can view this solution
as a singular gauge transformation 
\be 
\psi_{\phi}=U_\phi Q
U_\phi^{-1},\label{gaugesol} 
\ee 
where 
\be
U_\phi=1-\frac1{K+\phi}\phi Bc,\quad\quad
U_\phi^{-1}&=&1+\frac1{K}\phi Bc.\label{gaugetransf} 
\ee 
In order to prove that (\ref{psiphi}) is a solution, one demands that
$(c\phi)^2=0$, which requires the OPE of  $\phi$ at nearby points to
be not too singular.

It is instructive to write down the kinetic operator around
(\ref{psiphi}). With some manipulation, using the $K,B,c,\phi$
algebra one can show that \be {\cal Q}_{\psi_\phi} \frac
B{K+\phi}=Q\frac B{K+\phi}+\left\{\psi_\phi,\frac
B{K+\phi}\right\}=1.\0 \ee So, unless the homotopy--field $\frac
B{K+\phi}$ is singular, the solution has trivial cohomology, which
is the defining property of the tachyon vacuum
\cite{Ellwood,EllwoodSchnabl}. On the other hand, in order for the
solution to be well defined, the quantity
$\frac1{K+\phi}(\phi-\delta\phi)$ should be well defined. Moreover,
in order to be able to show that (\ref{psiphi}) satisfies the
equation of motion, one needs $K+\phi$ to be invertible.

In full generality we thus have a new nontrivial solution if
\begin{enumerate}
\item $\frac1{K+\phi}$ is singular, but
\item $\frac1{K+\phi}(\phi-\delta\phi)$ is regular and
\item $\frac1{K+\phi}(K+\phi)=1.$
\end{enumerate}
Problems with the last equation and Schwinger representation are discussed in appendix D.

In \cite{BMT} some sufficient conditions for $\phi$ to comply with
the above requirements were determined. Let us parametrize the
worldsheet RG flow, referred to above, by a parameter $u$, where
$u=0$ represents the UV and $u=\infty$ the IR, and rewrite $\phi$ as
$\phi_u$, with $\phi_{u=0}=0$. Then we require for $\phi_u$ the
following properties under the coordinate rescaling $f_t(z)=\frac
zt$ \be f_t\circ\phi_u(z)=\frac1t\,\phi_{tu}\left(\frac
zt\right)\label{cnd1} \ee and, most important, that the partition
function \be g(u)\equiv Tr[e^{-(K+\phi_u)}]=\left\langle
e^{-\int_0^1ds\, \phi_{u}(s) }\right\rangle_{C_1},\label{g(u)Tr} \ee
satisfies the asymptotic finiteness condition \be
\lim_{u\to\infty}\left\langle e^{-\int_0^1ds\, \phi_u(s)
}\right\rangle_{C_1}=\verb"finite".\label{cnd3} \ee Barring
subtleties, this satisfies the first two conditions above i.e.
guarantees not only the regularity of the solution but also its
'non-triviality', in the sense that if this condition is satisfied,
it cannot fall in the same class as the ES tachyon vacuum solution.
It would seem that the last condition above cannot be satisfied in
view of the first. But this is not the case. We will argue  in Appwndix D 
that by suitably defining the objects involved the
equation can indeed be satisfied.

We will consider in the sequel a specific relevant operator $\phi_u$
and the corresponding SFT solution. This operator generates an exact
RG flow studied by Witten in \cite{Witten}{},  see also
\cite{Kutasov}{}, and is based on the operator (defined in the
cylinder $C_T$ of width $T$ in the arctan frame)  
\be 
\phi_u(s) = u(X^2(s)+2\ln u +2 A)\label{TuCT1}, 
\ee 
where $A$ is a constant first
introduced in \cite{Ellwood}. In $C_1$  we have \be \phi_u(s) = u
(X^2(s)+ 2 \ln Tu +2 A)\label{TuC1b1} \ee and on the unit disk $D$,
\be \phi_u(\theta) = u (X^2(\theta)+ 2\ln \frac {Tu}{2\pi} +2
A).\label{TuDb1} \ee

If we set 
\be g_{A}(u)= \langle e^{-\int_0^1ds \,
\phi_u(s)}\rangle_{C_1} \label{gAu1} \ee we have \be g_{A}(u)
=\langle e^{- \frac 1{2\pi } \int_{0}^{2{\pi}} d\theta \, u\Bigl{(}
X^2(\theta) + 2 \ln \frac u{2\pi}+2 A\Bigr{)}}\rangle_{D}.  \0 \ee
According to \cite{Witten}, \be g_{A}(u) = Z(2u)e^{-2u (\ln \frac
u{2\pi}+A)},\label{gAu2} \ee where \be Z(u)= \frac 1{\sqrt{2\pi}}
\sqrt{u} \Gamma(u)e^{\gamma u}\label{Z(u)} \ee

Requiring finiteness for $u\to\infty$ we get $A= \gamma -1+\ln
4\pi$, which implies \be g_{A}(u)\equiv g(u)= \frac 1{\sqrt{2\pi}}
\sqrt{2u} \Gamma(2u) e^{2u(1-\ln (2u))}\label{gA1unorm} \ee and \be
\lim_{u\to\infty} g(u) = 1.\label{inflimitgA1u} \ee Moreover, as it
turns out, $\delta \phi_u=- 2u$, and so: \be \phi_u -\delta\phi_u =
u \d_u \phi_u(s).\label{uduphiu1} \ee Therefore the $\phi_u$ just
introduced satisfies all the requested properties and consequently
$\psi_u\equiv \psi_{\phi_u}$ must represent a D24 brane solution.

In I it was proved that $\psi_u$ can satisfy the closed string overlap condition
and it was seen that the corresponding RG flow in BCFT reproduces the correct ratio of
 tension between D25 and D24 branes. In I the  energy functional in SFT was also computed.

Let us now summarize the content of the paper.
In section 2 we write down the energy functional of the solution $\psi_u$ in the most convenient form
for the calculation, by isolating the 'angular' integration variables.
In section 3 we perform the integration over the latter, after which we are
left with two infinite discrete summations and an integral from 0 to $\infty$ over the
parameter $U$ (an alias of the RG parameter in CFT). We next carry out analytically one of
the discrete summations. The rest of the calculation we have been able to do only numerically.
In section 4 we analyze the behaviour near $U=0$ and describe our subtraction scheme for the UV singularity. In section 5 we study the behaviour as $U\to \infty$ and, with the help of some heuristics, we conclude that the energy integral converges in that region.
In section 6, we carry out the numerical evaluation of the energy functional.
In section 7 we introduce the regularized solution $\psi_\e$ and in section 8
we proceed to the evaluation of its energy, which turns out to vanish after subtracting
the UV singularity. In section 9 finally we compute the energy of $\psi_u-\psi_\e$ and find the desired result.

\section{The energy functional}

In I the expression for the energy of the lump solution was determined by evaluating a three--point
function on the cylinder $C_T$ of circumference $T$ in the arctan frame. It equals $-\frac 16$
times the following expression
\be
\langle \psi_u \psi_u\psi_u\rangle &=&-\int_{0}^\infty
dt_1dt_2dt_3{\cal
E}_0(t_1,t_2,t_3)u^3g(uT)\Bigg\{8\Big(-\frac{\partial_{2uT}g(uT)}{g(uT)}\Big)^3\0\\
&+&4\Big(-\frac{\partial_{2uT}g(uT)}{g(uT)}\Big)\Big(G_{2uT}^2(\frac{2\pi
t_1}T)+G_{2uT}^2(\frac{2\pi (t_1+t_2)}T)+G_{2uT}^2(\frac{2\pi
t_2}T)\Big)\0\\
&+&8G_{2uT}(\frac{2\pi t_1}T)G_{2uT}(\frac{2\pi
(t_1+t_2)}T)G_{2uT}(\frac{2\pi t_2}T)\Bigg\}.\label{LLL} \ee 
where $T=t_1+t_2+t_3$. Here
$g(u)$ is given by \be g(u)=\frac 1{\sqrt{2\pi}} \sqrt{2u}
\Gamma(2u) e^{2u(1-\ln(2u))}\label{g(u)} 
\ee 
and represents the
partition function of the underlying matter CFT on the boundary of
the unit disk with suitable boundary conditions for $u\to \infty$,
which will be discussed further on. $G_u(\theta)$ represents the
correlator on the boundary, first determined by Witten,
\cite{Witten}: 
\be G_u(\theta)= \frac {1}{u} +2 \sum_{k=1}^\infty
\frac {\cos (k\theta)}{k+u} \label{Gutheta} \ee where we have made
the choice $\alpha'=1$. Finally ${\cal E}_0(t_1,t_2,t_3)$ represents
the ghost three--point function in $C_T$. \be
\E_0(t_1,t_2,t_3)=\left\langle Bc\d c(t_1+t_2)\d c(t_1) \d c(0)
\right\rangle_{C_T} = -\frac 4{\pi} \sin \frac {\pi t_1}T \sin \frac
{\pi(t_1+t_2)}T \sin \frac{\pi t_2}T.\label{E0} \ee We change
variables $(t_1,t_2,t_3)\to (T,x,y)$, where \be x=\frac{t_1}{T},
\quad\quad y=\frac{t_2}{T}.\0 
\ee 
Then the matter part of
(\ref{LLL}) (before integration) can be written as $u^3 F(uT,x,y),$
where \be
F(uT,x,y)=&g(uT)\Bigg\{8\Big(-\frac{\partial_{2uT}g(uT)}{g(uT)}\Big)^3
+8G_{2uT}(2\pi x)G_{2uT}(2\pi(x+y))G_{2uT}(2\pi y)\0\\
+&4\Big(-\frac{\partial_{2uT}g(uT)}{g(uT)}\Big)\Big(G_{2uT}^2(2\pi
x)+G_{2uT}^2(2\pi(x+y))+G_{2uT}^2(2\pi y)\Big) \Bigg\}.\0
\ee
while the ghost correlator becomes
\be
\E_0(t_1,t_2,t_3)\equiv \E(x,y)=-\frac4\pi \,\sin\pi x\,\sin\pi y\,
\sin\pi(x+y).\label{Exy}
\ee
The ghost correlator only depends on $x$ and $y$, which are scale invariant coordinates.

After the change
\be
\int_0^\infty
dt_1 \int_0^\infty dt_2\int_0^\infty dt_3=\int_0^\infty dT\;
T^2\int_0^1 dx\int_0^{1-x} dy,\0
\ee
the energy becomes
\be
E[\psi_u]&=&-S[\psi_u]=-\frac 16\langle\psi_u\psi_u\psi_u\rangle\0\\
&=&\frac16\int_0^\infty dT\; T^2\int_0^1
dx\int_0^{1-x} dy\,\E(x,y)\,u^3 F(uT,x,y).\label{energy}
\ee
It is convenient to change further $x\to y$ and subsequently $y\to 1-y$. The result is
\be
E[\psi_u]&=\frac16\int_0^\infty dT\; T^2\int_0^1
dy\int_0^{y} dx\,\E(1-y,x)\,u^3 F(uT,1-y,x),\label{energy1}
\ee
where
\be
\E(1-y,x)=\frac4\pi \,\sin\pi x\,\sin\pi y\,\sin\pi(x-y)\0,
\ee
and
\be
&&F(uT,1-y,x)\label{F}\\
&&=g(uT)\Bigg\{8\Big(-\frac{\partial_{2uT}g(uT)}{g(uT)}\Big)^3
+8G_{2uT}(2\pi x)G_{2uT}(2\pi(x-y))G_{2uT}(2\pi y)\0\\
&&+4\Big(-\frac{\partial_{2uT}g(uT)}{g(uT)}\Big)\Big(G_{2uT}^2(2\pi
x)+G_{2uT}^2(2\pi(x-y))+G_{2uT}^2(2\pi y)\Big) \Bigg\}.\0
\ee

Summarizing
\be
E[\psi_u]&=&\frac16 \int_0^\infty d(2uT)\; (2uT)^2\int_0^1
dy\int_0^{y} dx\,\frac4\pi \,\sin\pi x\,\sin\pi y\,\sin\pi(x-y)\label{Efinal}\\
&&\cdot g(uT)\Bigg\{-\Big(\frac{\partial_{2uT}g(uT)}{g(uT)}\Big)^3
+G_{2uT}(2\pi x)G_{2uT}(2\pi(x-y))G_{2uT}(2\pi y)\0\\
&&-\frac 12 \Big(\frac{\partial_{2uT}g(uT)}{g(uT)}\Big)\Big(G_{2uT}^2(2\pi
x)+G_{2uT}^2(2\pi(x-y))+G_{2uT}^2(2\pi y)\Big) \Bigg\}.\0
\ee

As already stressed in I, the first important remark about this expression is its independence of $u$. In the original BCFT of Witten, $u$ was the RG coupling running from 0 (the UV) to $\infty$ (the IR). In SFT $u$ is simply a gauge parameter, with the exception of the extreme values $u=0$ and $u=\infty$.

The expression in (\ref{Efinal}) implies three continuous integrations and, in the most complicated case, three infinite discrete summations.
At the best of our ability and knowledge, all these operations cannot be done analytically. Therefore the obvious strategy to evaluate (\ref{Efinal}) is to push as far as possible the analytic computations and bring the integral to a form accessible to numerical evaluation. This is what we will do in the sequel.

\section{The angular integration}

The first step in the evaluation of (\ref{Efinal}) consists in performing the `angular' $x,y$ integration. This will be done analytically. Let us consider for definiteness the most complicated
term, the cubic one in $G_U$ (from now on for economy of notation let us set $U=2uT$).
We represent $G_U$ as the series (\ref{Gutheta}) and integrate term by term in $x$ and $y$.
All these integrations involve ordinary integrals which can be evaluated by using standard tables,
or, more comfortably, Mathematica. It is a lucky coincidence that most integrals are nonvanishing only for specific values of the integers $k$.
We have, for instance,
\be
&&\int_0^1dy \int_0^ydx \sin(\pi x)\, \sin(\pi y)\,\sin (\pi(x-y)) \,\cos (2\pi k x)=
\frac 1{8\pi (k^2-1)},\quad k\neq 1\0\\
&&\int_0^1dy \int_0^ydx \sin(\pi x)\, \sin(\pi y)\,\sin (\pi(x-y)) \,\cos (2\pi x)=
\frac 3{32\pi}\label{int1}
\ee
while the integral $\int_0^1dy \int_0^ydx \sin(\pi x)\, \sin(\pi y)\,\sin (\pi(x-y)) \,\cos (2\pi k x) \cos(2\pi m y)$ vanishes for almost all $k,m$ except $k=m,m\pm 1$ and $k=1,m$ and $k,m=1$.
For example
\be
&&\int_0^1dy \int_0^ydx \sin(\pi x)\, \sin(\pi y)\,\sin (\pi(x-y))\,\cos (2\pi k x)\cos(2\pi k y)= \frac 1{16\pi (k^2-1)}\label{int2}\\
&&\int_0^1dy \int_0^ydx \sin(\pi x)\, \sin(\pi y)\,\sin (\pi(x-y)) \cos (2\pi k x)
\cos(2\pi (k+1) y)= -\frac 1{32\pi k(k+1)},\quad\quad {\rm etc.}\0
\ee

The integration with three cosines is of course more complicated, but it can nevertheless be done in all cases. The integrals mostly vanish except for specific values of the integers $k,m,n$ inside the cosines. They are non-vanishing for $m=k$ with $n$ generic, and $m=k,n=k,k\pm 1$:
\be
&&\int_0^1dy \int_0^ydx \sin(\pi x)\, \sin(\pi y)\,\sin (\pi(x-y)) \label{int3}\\
&&\quad\quad\cdot\cos (2\pi k x)
\cos(2\pi k y)\cos(2\pi n(x- y))= \frac {n^2+k^2-1}{16\pi ((n+k)^2-1)((n-k)^2-1)}\0\\
&&\int_0^1dy \int_0^ydx \sin(\pi x)\, \sin(\pi y)\,\sin (\pi(x-y)) \0\\
&&\quad\quad\cdot\cos (2\pi k x)
\cos(2\pi k y)\cos(2\pi k(x- y))=- \frac {3(2k^2-1)}{16\pi (4k^2-1)}\0\\
&&\int_0^1dy \int_0^ydx \sin(\pi x)\, \sin(\pi y)\,\sin (\pi(x-y)) \0\\
&&\quad\quad\cdot\cos (2\pi k x)
\cos(2\pi k y)\cos(2\pi (k+1)(x- y))= \frac {6k^3+9k^2+3k-1}{128\pi (2k+1)(k+1)k}\0\\
&&\int_0^1dy \int_0^ydx \sin(\pi x)\, \sin(\pi y)\,\sin (\pi(x-y)) \0\\
&&\quad\quad\cdot\cos (2\pi k x)
\cos(2\pi k y)\cos(2\pi (k-1)(x- y))= \frac {6k^3-9k^2+3k+1}{128\pi (2k-1)(k-1)k}\0
\ee
and so on. A delicate part of the program consists in finding all nonvanishing terms and identifying the nonoverlapping ranges of summation over $k,m$ and $n$. Fortunately the  triple infinite summation reduces to a finite number of double infinite summations. Mathematica knows how to do the summations over one discrete index, in general not over two.

Let us write down next the result of the angular integration, by considering the three different
types of terms in (\ref{Efinal}) in turn.

\subsection{The term without $G_U$}

This is easy. We get
\be
&&\frac16 \int_0^\infty dU\; U^2\int_0^1
dy\int_0^{y} dx\,\frac4\pi \,\sin\pi x\,\sin\pi y\,\sin\pi(x-y)
 g(U)\Bigg[-\Big(\frac{\partial_{U}g(U)}{g(U)}\Big)^3\Bigg]\label{Efinal1}\\
&=&-\frac 1{4\pi^2}  \int_0^\infty dU\; U^2g(U)\Bigg[-\Big(\frac{\partial_{U}g(U)}{g(U)}\Big)^3\Bigg].\0
\ee

\subsection{The term quadratic in $G_U$}

We have to compute
\be
&&\frac16 \int_0^\infty d(U)\; (U)^2\int_0^1
dy\int_0^{y} dx\,\frac4\pi \,\sin\pi x\,\sin\pi y\,\sin\pi(x-y)\label{Efinal2}\\
&&\cdot (-\frac 12) \partial_{U}g(U)\Big(G_{U}^2(2\pi
x)+G_{U}^2(2\pi(x-y))+G_{U}^2(2\pi y)\Big) \Bigg\}.\0
\ee
Therefore the integrand of the quadratic term in $G_U$ is made of the factor
$-\frac 1{12} U^2 \partial_{U}g(U)$ multiplied by the factor
\be
&&\frac4\pi\int_0^1dy\int_0^ydx\, \sin\pi x\,\sin\pi y\,\sin\pi(x-y)\0\\
&&\quad\quad\cdot\Big(G_{U}^2(2\pi
x)+G_{U}^2(2\pi(x-y))+G_{U}^2(2\pi y)\Big) \Bigg\}.\label{GU2}
\ee
After some work the latter turns out to equal
\be
&&\frac4\pi\int_0^1dy\int_0^ydx\, \sin\pi x\,\sin\pi y\,\sin\pi(x-y)\0\\
&&\quad\quad\cdot\Big(G_{U}^2(2\pi
x)+G_{U}^2(2\pi(x-y))+G_{U}^2(2\pi y)\Big) \Bigg\}\label{GU21}\\
&=&-\frac 9{2\pi^2}\frac 1{U^2} + \frac {16}{\pi U} \left( \frac 9{32 \pi} \frac 1{U+1}+
\frac 3{8\pi} \sum_{k=2}^\infty \frac 1{k^2-1} \frac 1{k+U}\right)\0\\
&&+\frac {48}{\pi}\Biggl{(} \frac 1{8\pi} \sum_{\stackrel{k,n}{n\neq k,k\pm 1}}^{\infty} \frac {n^2+k^2-1}{((n+k)^2-1)((n-k)^2-1)}\frac 1{(n+U)(k+U)}\0\\
&&-\frac 1{4\pi} \sum_{k=1}^{\infty}
\frac {3k^2-1}{4k^2-1} \frac 1{(k+U)^2}\quad\leftarrow R_1(U)\0\\
&& +\frac 1{64\pi} \sum_{k=1}^{\infty} \frac {3k^2+3k+1}{k(k+1)}\frac 1{(k+U)(k+U+1)}\quad\leftarrow R_2(U)\0\\
&&+\frac 1{64\pi} \sum_{k=2}^{\infty} \frac {3k^2-3k+1}{k(k-1)}\frac 1{(k+U)(k+U-1)}\quad\leftarrow R_3(U)\Biggr{)}\0\\
&\equiv& E_{1}^{(2)}(U) + \frac {48}{\pi}\Biggl{(}\sum_{p=2}^{\infty}  RK(p,U)+ R_1(U)+R_2(U)+R_3(U)\Biggr{)},\0
\ee
where
\be
E_{1}^{(2)}(U)= -\frac 9{2\pi^2}\frac 1{U^2} +E_{0}^{(2)}(U)\label{E0E1}
\ee
and
\be
E_{0}^{(2)}(U) =\frac 9{2 \pi^2} \frac 1{U(U+1)}+ \frac 3{2\pi^2}
\frac 1{U(U^2-1)}\biggl{(}3(U+1)-4\gamma -4\psi(2+U)\bigg),\label{E21}
\ee
where $\psi$ is the digamma function and $\gamma$ the Euler--Mascheroni constant.
To save space, we have introduced in (\ref{GU21}) in a quite unconventional way the definitions of the quantities $R_i(U)$, $i=1,2,3$. Beside $R_1(U),R_2(U),R_3(U)$, we define
\be
RK(k,n,U)= \frac 1{8\pi} \frac {n^2+k^2-1}{((n+k)^2-1)((n-k)^2-1)}\frac 1{(n+U)(k+U)}\label{RK}
\ee
and
\be
RK(p,U)=\sum_{k=1}^\infty RK(k,k+p,U)+ \sum_{k=p+1}^\infty RK(k,k-p,U)\label{RKpm}
\ee
with the summation in (\ref{GU21}) corresponding to:
$\sum_{\stackrel{k,n}{n\neq k,k\pm 1}}^{\infty}RK(k,n,U)= \sum_{p=2}^\infty RK(p,U)$.

\subsubsection{Performing one discrete summation}

As already pointed out it is possible to perform in an analytic way
at least one of  the two discrete summations above. To start with
\be
R(U) &=& R_1(U)+R_2(U)+R_3(U)=  \frac{1}{32 \pi }\Bigl(\frac{U^2 (1+3 U)-2(U+1) H(U)}{U(1+U) (U^2-1)}\label{RU1}\\
&&+\frac{4 \left(1+4 U (1-\gamma+U-\ln 4)-4 U \psi(1+U)-2 \left(1-7
U^2+12 U^4\right) \psi^{(1)}(1+U)\right)} {\left(1-4
U^2\right)^2}\Bigr).\0 \ee Next \be
&&RK(p,U)= \frac 1{4 p \left(-1+p^2\right) \pi  (-1+p-2 U) (1+p-2 U) (-1+p+2 U) (1+p+2 U)}\0\\
&&\cdot\Biggl(4 p \left(-1+p^2\right) U \, H\left(\frac {p-1}2\right)\0\\
&&-(-1+p+2 U) (1+p+2 U) \left(-1+p^2-2 p U+2 U^2\right) H(U)\0\\
&&+(-1+p-2 U) \label{RKpU1}\\
&&\cdot\biggl(-(-1+p) p (1+p+2 U)+(1+p-2 U) \left(-1+p^2+2 p U+2
U^2\right) H(p+U)\biggr)\Biggr),\0 \ee where $H(U)= \gamma+
\psi(U+1)$ is the harmonic number function. It should be remarked
that in both (\ref{RU1}) and (\ref{RKpU1}) there are zeros in the
denominators, for positive values of U. These however do not
correspond to real poles of $R(U)$ and $RK(p,U)$, because they are
cancelled by corresponding zeroes in the numerator.

\subsection{The term cubic in $G_U$}

In (\ref{Efinal}) we have to compute
\be
\frac2{3\pi} \int_0^\infty dU\, U^2\int_0^1
dy\int_0^{y} dx\,\sin\pi x\,\sin\pi y\,\sin\pi(x-y)\, g(U) G_{U}(2\pi x)G_{U}(2\pi(x-y))G_{U}(2\pi y).\0\\
\label{Efinal3}
\ee
The most convenient form of the cubic term in $G_U$ after angular integration is probably the following one (which must be multiplied by $\frac 16 U^2g(U)$)
\be
&&\frac 4{\pi}\int_0^1dy\int_0^y dx \,\sin(\pi x)\, \sin(\pi y)\,\sin (\pi(x-y))\left(\frac 1U + 2\sum_{k=1}^\infty \frac {\cos(2\pi k x)}{k+U} \right)\label{int3GU}\\
&&\cdot \left(\frac 1U + 2\sum_{m=1}^\infty \frac {\cos(2\pi m y)}{m+U} \right)\left(\frac 1U + 2\sum_{n=1}^\infty \frac {\cos(2\pi n (x-y))}{k+U} \right)\0\\
&=&-\frac 3{2\pi^2}\frac 1{U^3} + \frac 9{4\pi^2} \frac 1{U^2(U+1)}+\frac 3{\pi^2} \frac 1{U^2(U^2-1)}\left(-\gamma +\frac 34 (U+1) -\psi(2+U) \right)\0\\
&&+\frac 3{4\pi^2}\frac 1{U(U+1)^2}-\frac 7{2\pi^2}\frac 1{U(U+1)(U+2)}+\frac 3{4\pi^2}\frac 1{U(U^2-1)^2}\cdot\0\\
&& \cdot \Bigl{(}3(1+U^2)-8\gamma U+6U -8U\psi(2+U)+4(U^2-1)\psi^{(1)}(2+U)\Bigr{)}\0\\
&&-\frac 1{2\pi^2U(U+1)(U^2-1)}\Bigl{(}17+5U-12\gamma-12\psi(3+U) \Bigr{)}- \frac 3{2\pi^2U^2(U^2-1)}\0\\
&&\cdot \Bigl{(}5-4\gamma +U -2(U+1)\psi(2+U) +2(U-1) \psi(3+U)\Bigr{)}\0\\
&&+\frac {32}{\pi}\,\Biggl[\, 3\sum_{\stackrel{k,n}{n\neq k,k\pm 1}} \frac {n^2+k^2-1}{16 \pi ((n+k)^2-1)((n-k)^2-1)}\, \frac 1{(k+U)^2(n+U)}\leftarrow (S_4,S_5)\0\\
&& -\,3\sum_{\stackrel{k,n}{n\neq k,k\pm 1,k-2}} \frac {n^2+k^2-k}{32 \pi (k^2-n^2)((k-1)^2-n^2)} \,\frac 1{(k+U)(k+U-1)(n+U)}\leftarrow (S_8)\0\\
&&-\,3\sum_{\stackrel{k,n}{n\neq k,k\pm 1,k+2}}\frac {n^2+k^2+k}{32 \pi (k^2-n^2)((k+1)^2-n^2)} \,\frac 1{(k+U)(k+U+1)(n+U)}\leftarrow (S_{9})\0\\
&&- 3 \sum_{k=1}\frac {2k^2-1}{16 \pi (4k^2-1)}\, \frac 1{(k+U)^3}\leftarrow (S_{10})\0\\
&&+ 2\sum _{k=1} \frac {6k^3+9k^2+3k-1}{128\pi k(k+1)(2k+1)} \,\frac 1{(k+U)^2(k+U+1)}\leftarrow (S_7)\0\\
&& +2 \sum _{k=2} \frac {6k^3-9k^2+3k+1}{128\pi k(k-1)(2k-1)} \,\frac 1{(k+U)^2(k+U-1)}\leftarrow (S_6)\0\\
&&-2\sum_{k=3}\frac {4k^2-8k+5}{64\pi(2k-1)(2k-3)}\, \frac 1{(k+U)(k+U-1)(k+U-2)}\leftarrow (S_2)\0\\
&& + \sum _{k=2} \frac {6k^3-9k^2+3k-1}{128\pi k(k-1)(2k-1)} \,\frac 1{(k+U)(k+U-1)^2}\leftarrow (S_{11})\0\\
&&-2\sum_{k=2}\frac {4k^2+1}{64\pi (4k^2-1)}\, \frac 1{(k+U)(k+U-1)(k+U+1)}\leftarrow (S_1)\0\\
&&+\sum _{k=1} \frac {6k^3+9k^2+3k+1}{128\pi k(k+1)(2k+1)} \,\frac 1{(k+U)(k+U+1)^2}\leftarrow (S_{12})\0\\
&&-2\,\sum_{k=1}\frac {4k^2+8k+5}{64\pi(2k+1)(2k+3)}\, \frac 1{(k+U)(k+U+1)(k+U+2)}\leftarrow (S_3)\Biggr].\0
\ee
The symbols $S_i$, $i=1,\ldots,12$ represents the corresponding terms shown in the formula and correspond to simple summations. As $S_4,S_5,S_8,S_9$ are shown in correspondence with double summations, they need a more accurate definitions. $S_4$ is the sum over $k$  from 2 to $\infty$ of the corresponding term for $n=k+2$, while $S_5$ is the sum of the same term from 3 to $\infty$ for $n=k-2$; $S_8$ is the sum over $k$ from 2 to $\infty$ of the corresponding term for $n=k+2$. $S_9$ is the sum over $k$ from 3 to $\infty$ of the corresponding term for $n=k-2$.

The first line of the RHS refers to the terms with one cosine, the next four lines to terms with 2 cosines and the remaining ones to terms with three cosines integrated over. In (\ref{int3GU}), $\psi^{(n)}$ is the $n$-th polygamma function and $\psi^{(0)}=\psi$.
There are simple and quadratic poles at $U=1$, but they are compensated by corresponding zeroes
in the numerators. One can also see that all the summations are (absolutely) convergent for any finite $U$, including $U=0$.

To proceed further let us define
\be
SK0(k,n,U)&=& \frac {n^2+k^2-1}{16 \pi ((n+k)^2-1)((n-k)^2-1)}\, \frac 1{(k+U)^2(n+U)}\0\\
SK1(k,n,U)&=& \frac {n^2+k^2-k}{32 \pi (k^2-n^2)((k-1)^2-n^2)} \,\frac 1{(k+U)(k+U-1)(n+U)}\0\\
SK2(k,n,U)&=& \frac {n^2+k^2+k}{32 \pi (k^2-n^2)((k+1)^2-n^2)} \,\frac 1{(k+U)(k+U+1)(n+U)}\0
\ee
and set
\be
SK0_+(p,U)&=& \sum_{k=1}^{\infty} SK0(k,k+p,U),\quad\quad
SK0_-(p,U)= \sum_{k=p+1}^{\infty} SK0(k,k-p,U),\0\\
SK1_+(p,U)&=& \sum_{k=2}^{\infty} SK1(k,k+p,U),\quad\quad
SK1_-(p,U)= \sum_{k=p+1}^{\infty} SK1(k,k-p,U),\0\\
SK2_+(p,U)&=& \sum_{k=1}^{\infty} SK2(k,k+p,U),\quad\quad
SK2_-(p,U)= \sum_{k=p+1}^{\infty} SK2(k,k-p,U).\0
\ee
Then the quantity within the square brackets in (\ref{int3GU}) corresponds to
\be
&&3\sum_{p=3}^\infty \biggl( SK0_+(p,U)+SK0_-(p,U) - SK1_+(p,U)-SK1_-(p,U)\0\\
&&\quad\quad -SK2_+(p,U)-SK2_-(p,U)\biggr)+
\sum_{i=1}^{12} S_i(U). \label{totalsumofS}
\ee

\subsubsection{Performing one discrete summation}

Like in the quadratic term we can carry out in an analytic way one discrete summation.
We have
\be
&&S(U)\label{SU}\\
&=&\sum_{i=1}^{12} S_i(U) =\frac 1{256 \pi  (-2+U) U^2 \left(-1-U+4 U^2+4 U^3\right)^3 \left(18-9 U-17 U^2+4 U^3+4 U^4\right)^2}\0\\
&&\cdot \Bigg(\frac{1}{3+U}
\biggl(12 \gamma (1+U) (2+U) (3+U)\0\\
&&\cdot \left(-324+13887 U^2-48589 U^4+72468 U^6-44592 U^8+11200 U^{10}\right)\0\\
&&+U^2 \bigl(
 (1+2 U)^3  (-845856+1192986 U+1878099 U^2-2889638 U^3-2109474 U^4\0\\
 &&+3023246 U^5
+1453619 U^6-1668346 U^7-622980 U^8+493352 U^9\0\\
&&+147696 U^{10}
-82016 U^{11}-21440 U^{12}+6016 U^{13}+1536 U^{14})
  \0\\
&&-192 (-2+U) (1+U)^3 (3+U) \left(-2+U+U^2\right)^2\0\\
&&\cdot \left(153-132 U^2+112 U^4+64 U^6\right) \ln 4\bigr)\Bigr)\0\\
&&+12 (1+U) (2+U) \0\\
&&\cdot\Bigl(\left(-324+13887 U^2-48589 U^4+72468 U^6-44592 U^8+11200 U^{10}\right) \psi(1+U)\0\\
&&+U \Bigl((-2+U) (-1+U) (1+U) (2+U) (-3+2 U) (-1+2 U) (1+2 U) (3+2 U) \0\\
&&\cdot\left(9+138 U^2-352 U^4+160 U^6\right) \psi^{(1)}(1+U)\0\\
&&+2 U \left(4-9 U^2+2 U^4\right) \left(-9+U^2
\left(7-4 U^2\right)^2\right)^2 \psi^{(2)}(1+U)\Bigr)\biggr)\Biggr).
\ee
Similarly
\be
&&SK(p,U)\label{SKpU}\\
&&\equiv \sum_{i=0}^2SKi_+(p,U)+SKi_-(p,U)  \0\\
&&=\frac{1}{32 p^2 \pi }\Biggl(\frac{2 p}{(1+p) (1+p-2 U)^2}+\frac{2 p}{(1+p)^2 (1+p-2 U)}+\frac{p^2}{\left(2+3 p+p^2\right) (2+p-2 U)}\0\\
&&+\frac{-1+p-p^2-2 p^3}{(-1+p)^2 \left(2+3 p+p^2\right) (1+U)}+\frac{4+p (-3+p (2+(-2+p) p))}{(-2+p) \left(-1+p^2\right)^2 (p+U)}+\frac{1+p (3+p)}{(1+p)^2 (2+p) (1+p+U)}\0\\
&& +\frac{p}{(-1+p) (2-p+2 U)}-\frac{2 p}{(1+p)^2 (-1+p+2 U)}+\frac{2 \left(-2+p^2\right)}{\left(-1+p^2\right) (p+2 U)}\0\\
&&-\frac{2 p}{(-1+p) (1+p+2 U)^2}+\frac{4 p}{(-1+p)^2 (1+p) (1+p+2 U)}
-\frac{2 (-1+p) p}{(-2+p) (1+p) (2+p+2 U)}\0
\ee
\be
&&+2 p \biggl(-\frac{8 \left(p-2 p^3+p^5-16 p U^4\right) \psi(\frac{1+p}{2})}{\left(\left(-1+p^2\right)^2-8 \left(1+p^2\right) U^2+16 U^4\right)^2}\0\\
&&+\frac{8 p \left(-4+p^2+4 U^2\right) \psi(\frac{2+p}{2})}{(-2+p-2 U) (p-2 U) (2+p-2 U) (-2+p+2 U) (p+2 U) (2+p+2 U)}\0\\
&&+\frac{1}{-1+p^2}\0\\
&&\cdot \Bigl(2\frac{2-4 p^4+21 p^3 U+6 U^2-8 U^4+p^2 \left(2-38 U^2\right)+p U \left(-9+28 U^2\right) \psi(1+U)}{(-2+p-2 U) (-1+p-2 U)^2 (p-2 U) (1+p-2 U)^2 (2+p-2 U)}\0\\
&&-2\frac{ \left(-2+4 p^4+21 p^3 U-6 U^2+8 U^4+p U \left(-9+28 U^2\right)+p^2 \left(-2+38 U^2\right)\right) \psi(p+U)}{(-2+p+2 U) (-1+p+2 U)^2 (p+2 U) (1+p+2 U)^2 (2+p+2 U)}\0\\
&&+\frac{\left(-1+p^2-2 p U+2 U^2\right) \psi^{(1)}(1+U)}{-1+p^2-4 p U+4 U^2}
-\frac{\left(-1+p^2+2 p U+2 U^2\right) \psi^{(1)}(1+p+U)}{-1+p^2+4 p U+4 U^2}\Bigr)\biggr)\Biggr).\0
\ee

\vskip 1cm

As explained above, in general we cannot proceed further with analytic means in performing
the remaining summations and integrations. The strategy from now on consists therefore in
making sure that summations and integrals converge (apart from the expected UV singularity,
which has to be subtracted). Let us study first the behaviour at $U\approx 0$. We will  proceed
next to the behaviour at $U\to \infty$.

\section{Behaviour near $U=0$}

Let us consider first the cubic term. We recall that all the summations are
convergent at $U=0$. In (\ref{Efinal3}) the expression (\ref{int3GU}) is multiplied
by $\frac 16 U^2g(U)$. Recalling that $g(U) \approx \frac 1{2\sqrt{\pi U}}$ for $U\approx 0$,
we see that the only term that produces a non-integrable singularity in $U$ is the first term
on the RHS, which has a cubic pole in $U$.  Altogether the UV singularity due to the cubic term is
\be
-\frac 18 \frac 1{\pi^{\frac 52} U^{\frac 32}}.\label{UV3}
\ee

As for the quadratic term, we have $\d_U g(U)\approx -\frac 1 {4\sqrt{\pi} U^{\frac 32}}$. Once again all the discrete summations are convergent at $U=0$. Therefore the only UV singular term corresponds to the first term at the RHS of (\ref{GU21}), i.e. $-\frac 9{2\pi^2}\frac 1{U^2}$.
According to (\ref{Efinal2}) we have to multiply this by $-\frac 1{12} U^2 \d_U g(U)$. Therefore
the contribution of the quadratic term to the UV singularity is
\be
-\frac 3{32} \frac 1{\pi^{\frac 52} U^{\frac 32}}.\label{UV2}
\ee

Finally for the last term, the one without $G_U$, we have
\be
U^2 g(U) \left(\frac {\d_U g(U)}{g(U)}\right)^3\approx
-\frac 1{16\sqrt{\pi} U^{\frac 32} } .\0
\ee
Therefore altogether this term contributes
\be
-\frac 1{64} \frac 1{\pi^{\frac 52} U^{\frac 32}}.\label{UV1}
\ee
So the overall singularity at $U=0$ is
\be
-\frac {15}{64} \int_0 dU \frac 1{\pi^{\frac 52} U^{\frac 32}}= \frac {15}8
\frac 1{4\pi^2\sqrt{\pi U}}\Biggr{|}_{U=0}= - \lim_{U\to 0}
 \frac {15}8 \frac 1{2\pi^2}\frac 1{2\sqrt{\pi U}}.\label{UV}
\ee

In order to subtract this singularity we choose a function
$f(U)$ that vanishes fast enough at infinity and such that $f(0)=1$.
For instance $f(U)= e^{-U}$. Then, if we subtract from the energy the expression
\be
&& \frac {15}8 \frac 1{4\pi^2\sqrt{\pi}} \int_0^{\infty}dU \frac 1{\sqrt{U}}\left(f'(U)
 - \frac 1{2U} f(U)\right)= \frac {15}8 \frac 1{4\pi^2\sqrt{\pi}}
  \int_0^{\infty}dU \frac {\d}{\d U} \left(\frac 1{\sqrt{U}}f(U)\right) \label{subtr}\0\\
\ee
the energy functional becomes finite, at least in the UV.  What remains after the subtraction
is the relevant energy.

Notice that the integral in (\ref{subtr}) does not depend on the regulator $f$ we use, provided it satisfies the boundary condition $f(0)=1$ and decreases fast enough at infinity.
As we shall see in section 9, the lump energy is anyhow thouroughly independ of such UV subtractions.

\section{The behaviour near $U=\infty$}

The integrand in (\ref{Efinal1}) behaves as $1/U^4$ at large $U$. Therefore the integral (\ref{Efinal1}) converges rapidly in the IR.

\subsection{The quadratic term as $U\to\infty$}

With reference to (\ref{Efinal2}) we remark first that for large $U$
\be
U^2\d_U g(U)= -\frac 1{12 \sqrt{2}}+{\cal O}\left(\frac 1U\right).\label{Dguasymp}
\ee
Therefore this factor does not affect the integrability at large $U$. The issue will be decided
by the other factors. For large $U$ we have
\be
E_{0}^{(2)}(U) = \frac 9{\pi^2} \frac 1{U^3}- \frac 6{\pi^2} \frac {\ln U}{U^4}+\cdots\label{E20}
\ee
and
\be
R(U)= -\frac 3{32\pi} \frac 1U + \frac 1{16\pi} \frac 1{U^2}-\frac 3{32\pi} \frac{\ln U}{U^3}+\cdots.\label{RUasym}
\ee
Moreover, again for large $U$,
\be
RK(p,U)= \frac 1{8\pi(p^2-1)} \frac 1U- \frac 1{8\pi(p^2-1)} \frac 1{U^2}-\frac 1{16\pi}
\frac{\ln U}{U^3}+\cdots.\label{RKUasym}
\ee
Since
\be
\sum_{p=2}^\infty \frac 1{8\pi(p^2-1)}=\frac 3{32\pi}\0
\ee
the coefficient of $1/U$ in (\ref{RUasym}) cancels the corresponding coefficient of (\ref{RKUasym}). The coefficient of $1/U^2$ equals $-1/(32\pi)$. This must be multiplied by $\frac {48}{\pi}$ and
added to the term $-\frac 9{2\pi^2}\frac 1{U^2}$ in (\ref{GU21}). This is anyhow an integrable term in the IR.
This much takes care of the integrability of the $E_{0}^{(2)}(U)$, $R(U)$ and the first two terms in (\ref{RKUasym}) in the IR. Let us now concentrate on the rest of $RK(p,U)$, that is
\be
RK'(p,U)= \frac 1{16\pi} \frac{\ln U}{U^3}+\ldots \label{RKcpUasym}
\ee
(see a more complete asymptotic expansion in Appendix A).

In order to estimate the integrability of this term, we can replace
the  infinite discrete sum with an integral over $p$, for large $p$.
Now we evaluate the behaviour of $RK'(p,U)$ for any ray, departing
from the origin of the $(p,U)$ plane in the positive quadrant, when
the rays approach infinity. We can parametrize a ray, for instance,
as the line $(a U,U)$.  It is possible to find an analytic
expression for this. We can compute the large $U$ limit for any
(positive) value of $a$. The behaviour is given by the following
rule \be RK'(a U,U)\approx \frac {B}{U^3}+ {\cal
O}(U^{-4}).\label{RUUasym} \ee In Table 1 are some examples (the
output is numerical only for economy of space). \TABLE[ht]{
$\begin{matrix}\quad &a: & 45\,&7\,&1 \, &  \frac 1{13}\,&\frac 1{150} & \frac 1{15000}\\
\quad&B:\quad&-0.00001\quad&-0.00048\quad&\-0.00713\quad&-0.04319\quad&-0.09039\quad&-0.18188
\end{matrix}$
\caption{Samples of $a$ and $B$ in eq.(\ref{RUUasym})}
}
It is important to remark that very small values of $a$ are likely not to give a reliable response in the table, because one is bound to come across to the forbidden value
$p=1$, which will give rise to an infinity (see (\ref{RKpm}).
Apart from this, on a large range the values of $RK'(a U,U)$ are bounded in $a$.

It is even possible to find an analytic expression of $B$ as a function of $a$ in the large $U$ limit.
We have
\be
B&=& \frac{1}{8 a^3 \left(-4+a^2\right)^2 \pi }\Big(-a \left(16-8 a+2 a^3+a^4+8 a^2 (-1+\log 2)\right)\0\\
&&\quad\quad +8 a^3  \log a+2 (-2+a)^2 \left(2+2 a+a^2\right)  \log(1+a)\Big)\label{Ba}
\ee
which is obviously integrable in the whole range of $a$. Since $dpdU$=$UdadU$, this confirms the integral behaviour
of $\sum_{p=2}^\infty RK'(p,U)$ with respect to the $U$ integration.

To study the integrability for large $U$ and large $p$ in a more systematic way,
we divide the positive quadrant of the $(p,U)$ plane in a large finite number $N$ of small angular wedges. We notice that Table 1  means that
$RK'(p,U)$ varies slowly in the angular direction -- it is actually approximately constant in that direction for large $p$ and $U$. Therefore it is easy to
integrate over such wedges from a large enough value of the radius $r=\sqrt{p^2+U^2}$ to infinity. The result of any such integration will be a finite number and a good approximation to the actual value (which can be improved at will). Their total summation will also be finite as a consequence of table 1, unless there are pathologies at the extremities. Looking at the asymptotic expansion for large $p$
\be
RK(p,U)= \frac 1{4\pi} \frac {\log p}{p^3} -  \frac 1{4\pi}  (1+\psi(1+U)) \frac 1 {p^3}+\ldots\label{RKpUpasymp}
\ee
and Table 1 we see that also the integration for the very last wedge,  $a$ large, will be finite. The contribution of the very first wedge is more problematic for the above explained reason and is deferred to Appendix A.

An additional support comes from a numerical analysis of $RK'(p,U)$.
It turns out that, for large  $U$, the leading coefficient of
$\sum_{p=2}^\infty RK'(p,U)$ is \be \sum_{p=2}^\infty RK'(p,U)
\approx \frac {-0.0344761...}{U^2}+\cdots.\label{leadingRK'} \ee
This can be rewritten in the (probably exact) analytic way \be
\sum_{p=2}^\infty RK'(p,U) = \left(\frac 3{32 \pi} - \frac 1 {4\pi}
\left(\gamma +\frac 13 \log 2\right)\right) \frac
1{U^2}+\cdots.\label{leadinganalytic} \ee

Finally, the numerical calculations of the next section further confirm our conclusion.

On the basis of that analysis and the above, we conclude that the quadratic term integrand in $U$, behaves in the IR in an integrable way, giving rise there to a finite contribution
to the energy.

\subsection{The cubic term as $U\to\infty$}

To start with let us recall that for large $U$ \be U^2 g(U)=
\frac{U^2}{\sqrt{2}}+ \frac U{12 \sqrt{2}}+\cdots.\label{U2gU} \ee
Looking at (\ref{Efinal3}) and (\ref{int3GU}), let us call \be
E^{(3)}_1(U)&=&-\frac 3{2\pi^2}\frac 1{U^3} + \frac 9{4\pi^2} \frac 1{U^2(U+1)}+\frac 3{\pi^2} \frac 1{U^2(U^2-1)}\left(-\gamma +\frac 34 (U+1) -\psi(2+U) \right)\0\\
&&+\frac 3{4\pi^2}\frac 1{U(U+1)^2}-\frac 7{2\pi^2}\frac 1{U(U+1)(U+2)}+\frac 3{4\pi^2}\frac 1{U(U^2-1)^2}\cdot\0\\
&& \cdot \Bigl{(}3(1+U^2)-8\gamma U+6U -8U\psi(2+U)+4(U^2-1)\psi^{(1)}(2+U)\Bigr{)}\0\\
&&-\frac 1{2\pi^2U(U+1)(U^2-1)}\Bigl{(}17+5U-12\gamma-12\psi(3+U) \Bigr{)}- \frac 3{2\pi^2U^2(U^2-1)}\0\\
&&\cdot \Bigl{(}5-4\gamma +U -2(U+1)\psi(2+U) +2(U-1) \psi(3+U)\Bigr{)}.\label{E1}
\ee
Then it is easy to prove that
\be
\lim_{U\to \infty} U^3 E^{(3)}_1(U) =-\frac 3{2\pi^2}\label{limit1}
\ee
that is, the nonvanishing, nonintegrable, contribution comes solely from the first term on the RHS
of (\ref{E1}).
Defining $E^{(3)}_0(U) = E^{(3)}_1(U)+ \frac 3{2\pi^2}\frac 1{U^3}$, one finds
\be
E^{(3)}_0(U)\approx \frac 3{\pi^2}\frac  {\ln U}{U^4}+ \frac{3\gamma}{\pi^4}\frac 1{U^4}+\cdots.\label{approx2}
\ee
This corresponds to an integrable singularity at infinity in the $U$-integration.
We expect the nonintegrable contribution coming from (\ref{limit1}) to be cancelled
by the three--cosine pieces. We will see that also the first terms in the RHS of (\ref{approx2})
gets cancelled.

Let us see the three-cosines pieces in eq.(\ref{int3GU}). The first contribution (\ref{SU}),
for large $U$ goes as follows
\be
S(U)= \frac 5{256\pi}\frac 1{U^3}-\frac {2+\ln 8}{32\pi} \frac 1{U^4}+\cdots.\label{SUasym}
\ee
The other contribution is given by $SK(p,U)$. We proceed as for $RK(p,U)$ above.
\be
 SK(p,U)=\frac{1}{32 \pi p(p+2)} \frac 1{U^3}
 - \frac {1-p(p+1)\left(H\left(\frac {p-1}2\right) - H\left(\frac {p}2\right)\right)}{32\pi p(p+1)}\frac 1{U^4}
+\cdots.\label{SKUasym}
 \ee
Let us consider the first term in the RHS, which, from (\ref{int3GU}), must be multiplied by 3. The sum over $p$ up to $\infty$  gives the following coefficient of $1/U^3$
\be
3 \sum_{p=3}^{\infty}\frac{1}{32 \pi p(p+2)} = \frac 7{256 \pi}.\0
\ee
This must be added to the analogous coefficient in the RHS of (\ref{SU}), yielding
a total coefficient of $\frac 3{64 \pi}$. In eq.(\ref{int3GU}) this is multiplied by
$\frac {32}{\pi}$, which gives $\frac 3{2\pi^2}$. This cancels exactly the RHS of (\ref{limit1}).
Therefore {\it in the integral (\ref{int3GU}) there are no contributions of order
$1/U^3$ for large $U$}.

As already remarked for the quadratic term, the above takes care of the nonintegrable asymptotic behaviour of (\ref{limit1},\ref{SUasym}), but it is not enough as far as
the $SK(p,U)$ is concerned. We will proceed in a way analogous to the quadratic term.
We will drop the first term in the RHS of (\ref{SKUasym}) (since we know how to deal exactly with the latter) and define
\be
 SK'(p,U)=SK(p,U)-\frac{1}{32 \pi p(p+2)} \frac 1{U^3} .\label{SKc}
 \ee
In order to estimate the integrability of this expression, we will replace,  for large $p$, the infinite discrete sum with an integral over $p$. Next we evaluate the behaviour of $SK(p,U)$ for any ray departing from the origin of the $(p,U)$ plane in the positive quadrant when the rays approach infinity, parametrizing a ray as the line $(a U,U)$, $a$ being some positive number. The behaviour is given in general by the following rule
\be
SK(a U,U)\approx \frac {B}{U^5}+\cdots.\label{SKUUasym}
\ee
In Table 2 are some examples (the output is numerical for economy of space):
\TABLE[ht]{
$\begin{matrix}\quad &a: & 45\,&\,7&1 \,&1/{15}\, & 1/{85} &1/{150} \\
\quad&B:&-4\times 10^{-6}&-0.00017& -0.00497& -0.13988&-0.83567&
-1.4822\end{matrix}$
\caption{Samples of $a$ and $B$ in eq.(\ref{RUUasym})}
}
Also in this case we warn that it does not make sense to probe extremely small values  of
$a$.

On the other extreme, large $p$ and fixed $U$, we have
\be
SK(p,U)= \frac{-1+(1+U) \psi^{(1)}(1+U)}{16 \pi  (1+U)} \frac 1{p^3} +
\frac {-3 - 4 U + 4 U (1 + U)  \psi^{(1)}(1+U)} {32 \pi (1 + U)}
 +\cdots.\0\\\label{SKpUpasymp}
\ee
This behaviour is of course integrable at $p=\infty$. One can also verify a behaviour in $p$ similar to (\ref{SKUUasym}) and compute a table like Table 2.

Next we study the problem of integrability for large $U$ and large $p$ following the same pattern as for the quadratic term. We divide the positive quadrant of the $(p,U)$ plane in a large finite number $N$ of small angle wedges. We notice that Table 1  means that
$U^2 SK'(p,U)$ varies slowly in the angular direction -- it is actually approximately constant in that direction for large $p$ and $U$, see (\ref{U2SKcradial}) below. Therefore it is easy to
integrate $U^2 SK'(p,U)$ over such wedges from a large enough value of the radius $r=\sqrt{p^2+U^2}$ to infinity. The result of any such integration will be a finite number, including the integration for the very last wedge,  $a$ very large. To estimate the effectiveness of this approach one should consider the first wedge, which is the most problematic in view of what has been remarked above. But this point is very technical and we decided to postpone it to Appendix B.

Additional evidence for convergence can be provided by a numerical analysis. One can see that the behaviour of $U^2SK'(p,U)$ for large $p$ and $U$ may be approximated by
by
\be
U^2 SK'(p,U)\sim \frac {\log r}{r^3}\label{U2SKcradial}
\ee
which is integrable. We can do better and compute, numerically, the asymptotic behaviour
\be
U^2\sum_{p=3}^\infty SK'(p,U)\approx -\frac{0.0092 Log(U)}{U^2}+\dots \label{asymSK'}
\ee
The numerical calculations of the next section also confirm this.
So we conclude that for the cubic term too, the integrand in $U$ behaves in the IR in an integrable way, giving rise there to a finite contribution to the energy.

{\it Finally, on the basis of the heuristic analysis of this section, we conclude that, once the UV singularity is suitably subtracted, the energy integral (\ref{Efinal}) is finite.}

\section{Numerical evaluation}

This section is devoted to the numerical evaluation of (\ref{Efinal}) using the results
of the previous sections.

The first step is subtracting the UV singularity. We have already illustrated the method in section 4. It remains for us to do it in concrete by choosing a regulator.
Since we are interested in enhancing as much as possible the numerical convergence we will choose the
following families of $f$'s
\be
f(v)=\left\{ \begin{matrix} e^{-\frac v{a^2-v^2}} \quad & 0\leq v\leq a\cr
0 \quad &v\geq a\end{matrix}\right.  \label{fv}
\ee
where $a$ is a positive number.
It equals 1 at $v=0$ and 0 at $v=a$. Therefore, for terms in the integrand of (\ref{Efinal}) that are singular in $v=0$, we will split
the integral in two parts: from 0 to $a$, and from $a$ to $\infty$. The part from 0 to $a$
will undergo the subtraction explained in sec. 4.

We have checked the regulator for several values of $a$, $a=0.01,0.5,1,2,10,100,...$. Changing
$a$ may affect the fourth digit of the results below, which is within the error bars of our
calculations. Therefore in the sequel we will make a favorable choice for the accuracy of the calculations: $a=1$.

Let us proceed to evaluate the three terms in turn.

\subsection{The cubic term}

In eq.(\ref{int3GU}) we have to pick out the term $-\frac 3{2\pi^2} \frac 1{U^3} $ and treat it
separately. Let us consider it first in the range $0\leq U\leq 1$ and subtract the UV divergence. In the range
$1\leq U<\infty$ instead, according to the discussion of the last section, we will combine it with the most divergent of the remaining terms. This will render the corresponding integrals convergent.

1) Let us start with the subtraction for $-\frac 3{2\pi^2} \frac 1{U^3} $. Proceeding as explained above the subtracted integrand (after multiplying by $\frac 16 U^2 g(U)$) is
\be
- \frac 1{4\pi^2}\left(\frac {g(U)}U -\frac 1{\sqrt{\pi U}}
 \frac{e^{\frac{U}{U^2-4}} \left(16+8 U-8 U^2+2 U^3+U^4\right)}{2 U \left(U^2-4\right)^2} \right).\label{1stint}
\ee
This, integrated from 0 to 1, gives $-0.0619767 $.

2) Now, let us consider the term (\ref{E1}). Leaving out the first term we get $E_0^{(3)}(U)$.
When multiplied by $\frac 16 U^2 g(U)$ the result has integrable singularity at $U=0$, therefore
it can be directly integrated from 0 to $\infty$. The result is $0.109048$.

3) Next we have the term $S(U)$. When multiplied by $\frac 16 U^2 g(U)$, it is non-integrable at $\infty$. Thus we split  $-\frac 3{2\pi^2} \frac 1{U^3} $  as
$-\frac {32}{\pi} \frac 5{256\pi}  \frac 1{U^3} - \frac {32} {\pi} \frac 7{256\pi}\frac 1{U^3}$. We add
the first addend to  $S(U)$ in the range $1\leq U<\infty$, so as to kill the singularity at infinity. Then we multiply the result by $\frac {32}{\pi}\frac 16 U^2 g(U)$. The overall result is integrable both in 0 and at $\infty$. Finally we integrate from 0 to 1 and from 1 to $\infty$ the corresponding unsubtracted and subtracted integrands. The result
is $- 0.0190537$.

4) Now we are left with the $SK(p,U)$ terms. This must be summed over $p$ from 3 to infinity.
After summation this term must be multiplied by $\frac {96}{\pi}\frac 16 U^2 g(U)$.
The result is integrable in the UV, but not in the IR. In fact we must subtract the other
piece of  $-\frac 3{2\pi^2} \frac 1{U^3} $, more precisely we should add
$ - \frac {32} {\pi} \frac 7{256\pi}\frac 1{U^3}$ to (\ref{SKpU}) in the range $1\leq U\leq \infty$. The best way to do it is to split the integration in the intervals (0,1)
and (1,$\infty$), and to subtract from (\ref{SKpU}) the term $\frac 1{32\pi p(p+2)} \frac 1{U^3}$.
At this point we proceed numerically with Mathematica, both for the summation over $p$ and the integration over $U$. The result is $-0.029204$, with possible errors at the fourth digit.

 According to the above, the cubic term's overall contribution to the energy is $-0.00118596$.

\subsection{The quadratic term}

1) Also in this case, looking at (\ref{GU2},\ref{GU21}), we treat separately the term
$-\frac 9{2\pi^2} \frac 1 {U^2}$. This term must be multiplied by $-\frac 1{12} U^2 \d_Ug(U)$.
We get as a result
\be
s(U)=-\frac{3 e^U e^{-(\frac 12 +U)\ln U}
   \Gamma(U) (-1 + 2U \ln U - 2U \psi(U))}{ 32 \pi^{\frac 52}}.\label{sU}
 \ee
The resulting term is regular in the IR but singular in the UV. We make the same subtraction as above and obtain
\be
{\mathfrak s}(U)= s(U) + \frac 3{16 \pi^2} \frac 1{\sqrt{\pi U}}\frac{e^{\frac{U}{U^2-4}}
\left(16+8 U-8 U^2+2 U^3+U^4\right)}{2 U \left(U^2-4\right)^2}.  \label{sUbf}
  \ee
It is easy to see that this is now integrable also in the UV. Integrating it between 0 and 1
and $s(U)$ between 1 and infinity one gets $ 0.0379954$.

2) Next comes the integration of the term containing $E_0^{(2)}(U)$, see (\ref{E0E1}) above.
This must be multiplied also by $-\frac 13 U^2 \d_Ug(U)$. The result is
a function regular both at 0 and $\infty$. One can safely integrate in this range and get
$ 0.0156618 $.

3) The next term is $R(U)$, (\ref{RU1}).
This behaves like $\frac 1U$ for large $U$, see (\ref{RUasym}). So we subtract the corresponding divergent term, knowing already that
it cancels against the analogous behaviour of the $RK$ piece (see also below).
Therefore we define
\be
R'(U)= R(U) +\frac 3{32 \pi} \frac 1U.\label{R'U}
\ee
This has the right behaviour in the IR, but not the UV. For multiplying by
$-\frac 1{12} U^2 \d_Ug(U)$ one gets an ultraviolet singularity. The way out is to limit
the subtraction (\ref{R'U}) to the range $(1,\infty)$. This can be done provided we do the same with the $RK$ term, see below. Finally, in the range (0,1) we will integrate the term containing $R(U)$ without correction, since it can safely be integrated there. In the range $(1,\infty)$ we will integrate the one containing $R'(U)$. The overall result is $ -0.00392332 $.

4) There remains the $RK(p,U)$ piece, see (\ref{RKpU1}).
Again we have to subtract the singularity at $\infty$ (knowing that it cancels against the previous one). So we define
\be
RK'(p,U)= RK(p,U) - 1/(8 \pi (p^2 - 1) U).\label{RKcpU}
\ee
However, when multiplying by  $-\frac 4{\pi} U^2 \d_Ug(U)$,  this introduces an UV singularity, so in accordance with the previous subtraction,
this subtraction has to be limited to the range (1,$\infty$). Consequently we have also to split the integration. Both integrals from 0 to 1 and from 1 to $\infty$ are well defined. The numerical evaluation gives  $0.000235065 $.

The overall contribution of the quadratic term is therefore $ 0.049969$.

\subsection{Last contribution}

The last one is easy to compute. The integrand is
\be
\frac 1{4\pi^2} U^2 g(U) \left( \frac {\d_U g(U)}{g(U)}\right)^3.\label{last}
\ee
This converges very rapidly in the IR. The only problem is with the usual singularity
in the UV, where (\ref{last}) behaves like
$- \frac 1{64 \pi^2} \frac 1{\sqrt{\pi}U^{\frac 32}} $. To this end we will add to (\ref{last}) the function
\be
\frac 1{32 \pi^2} \frac 1{\sqrt{\pi U}}\frac{e^{\frac{U}{U^2-4}}
\left(16+8 U-8 U^2+2 U^3+U^4\right)}{2 U \left(U^2-4\right)^2}.   \label{additionlast}
  \ee
The sum of the two is now well behaved and can be integrated from 0 to $\infty$. The result is $0.0206096$.

\subsection{Overall contribution}

In conclusion the total finite contribution to the energy is $0.0693926$.
\be
E^{(s)}[\psi_u]\approx 0.0693926, \label{finalresult}
\ee
where the superscript $^{(s)}$ means that we have subtracted away the UV singularity.
This has to be compared with the expected D24 brane tension
\be
T_{D24}= \frac 1{2\pi^2}\approx 0.0506606.\label{TD24}
\ee
This theoretical value is justified in Appendix C. The two values (\ref{finalresult}) and (\ref{TD24}) differ by about $27\%$.

\subsection{Error estimate}

All the numerical calculations of this paper have been carried out with Mathematica.
Mathematica can be very precise when performing numerical manipulations. However
in our case there are two main sources of error, beside
the subtraction of the infinite D25-brane factor and the precision of Mathematica. The first is the summation over
$p$ of $RK(p,U)$ and especially $SK(p,U)$. The precision of this summation is probably limited by the computer capacity and seem to affect up to the fourth digit in
item 4 of section 6.1 and especially 6.2. Another source of errors is the presence
of zeroes in the denominators of the expressions of $RK(p,U)$ and $SK(p,U)$. As we have explained above, they do not correspond to poles, because they are canceled by corresponding zeroes in the  numerators; but Mathematica, when operating numerically, is not always able (or we have not been able to use it properly) to smooth out the corresponding functions. This again may affect the
fourth digit of item 4 of section 6.1 and especially 6.2.

It is not easy to evaluate these sources of error. A certain number of trials suggest that a possible error of $1\%$ in the final figure (\ref{finalresult}) does not seem to be unreasonable. We shall see that actually
the numerical result (\ref{finalresult}) we have obtained is more precise than that.

However it is clear from now that $E[\psi_u]$ {\it  is not the lump energy} we are looking for.
This may be a bit disconcerting at first sight, because, after all, we have subtracted from the energy
the UV singularity, which corresponds to tachyon vacuum energy. However one must reflect on the circumstance
that this subtraction contains an element of arbitrarness. In fact the subtraction is purely {\it ad hoc}, it is a subtraction on the energy functional alone, not a subtraction made in the framework of a consistent scheme.
In order to make sure that our result is physical we have to render it independent of the subtraction scheme.
This is what we will do in the next three sections. 
$\psi_u$ is a (UV subtracted) solution to the SFT equation of motion {\it on the perturbative vacuum}; what we need is the solution corresponding to $\psi_u$ {\it on the tachyon condensation vacuum}.
As we shall see, the gap between (\ref{finalresult}) and (\ref{TD24}) is the right gap between the (subtracted) energy of $\psi_u$ and the energy of the lump above the tachyon condensation vacuum.

\section{The regularized solution}

The solution to the puzzle came to us in a rather indirect way,
from an early development of our research, when we thought a
regularization  of our solution was necessary. The Schwinger
representation we use in our determination of the energy
(\ref{Efinal}) looks, at a superficial inspection, singular and in need of a regularization. In fact
it is not, as we show in Appendix D.  Instead, what actually
happens when we regularize the Schwinger parametrization in $\psi_u$
is that we turn it into the tachyon vacuum solution. But this
results into a happy occurrence because it suggests the way to an
analytic determination of the lump energy.

On a general ground it would seem necessary to regularize expressions like $1/K, 1/(K+1), 1/(K+\phi)$. The reason is the following. $K$ is a vector in an infinite dimensional vector space. Therefore an expression
like $1/K$ does not even make sense without a suitable specification.
As for $1/(K+1)$, we can understand it as the power series expansion
\be
\frac 1{K+1}= 1-K+K^2-\dots, \0
\ee
but does the series converge? if it converges, in what topology should the convergence be understood? We consider two ways to answer such questions.

One way is to view $1/K$ as the action of the operator $1/K_1^L$ on the identity state $|I\rangle$. The operator $K_L^1$ is a hermitean operator. Its spectrum is necessarily real. It has been studied in the series of papers \cite{RSZ,BMST1,BMST2,BMT3} and its spectrum extends over the full real axis
\footnote{In  \cite{BMST1,BMST2,BMT3}, in the ghost case, additional points of the spectrum were found outside the real axis, but only because the matrices $G,A,B,C,D$ used to represent $K_1^L$ are not hermitean.}.
Therefore not even the operator expressions $(K^L_1)^{-1}, (K^L_1+1)^{-1}$ make sense.
However operator theory teaches us that we can write perfectly sensible expressions
$ (K^L_1+\epsilon)^{-1} $ and $ (K^L_1+1+\epsilon)^{-1}$, provided $\epsilon$ has a non-vanishing imaginary part. For instance, an expression like $(K^L_1+\epsilon)^{-1} $ is analytic in the
$\epsilon$ plane outside the real axis (see \cite{DS}). We will call it the operator regularization.

The case $1/(K+\phi)$ is discussed in Appendix D. It is very plausible that $\phi$ may in general play the role of a regulator. In any case one would not see any a priori
harm in representing $1/(K+\phi)$ as $ (K+\phi+\epsilon)^{-1} $ in the $\e\to 0$ limit.

Another way of giving a meaning to expressions like $1/K, 1/(K+1), 1/(K+\phi)$ is by means of a
Schwinger representation (an extended version of the Hille-Phillips-Yosida theorem, see \cite{DS}).
For instance
\be
\frac 1{K+1} = \int_0^{\infty} dt \, e^{-t(K+1)}\label{schwinger1}
\ee
is a well-known example of regular representation (although it is not known if it is regular
for all correlators). We will therefore give a meaning to such expressions as
$1/K, 1/(K+\phi)$ by means of the regularized Schwinger representations
\be
\frac 1{K+\e} = \int_0^{\infty} dt \, e^{-t(K+\e)},\quad\quad
\frac 1{K+\phi+\e} = \int_0^{\infty} dt \, e^{-t(K+\phi+\e)}\label{schwinger2}
\ee
in the limit $\e\to 0$.
We remark that the Schwinger regularization usually converges for real $\epsilon>0$, while
the operator regularization requires $\Im (\epsilon) \neq 0$.  

So far our argument has been classical (in the sense of classical operator theory) and one would not expect any harm from such regularizations, but in fact they are not innocuous in the quantum theory.
From now on, in this section and the next, we will use the above regularized Schwinger representations and $\epsilon$ will be our regulator.
We will show that the introduction of this innocent looking $\e$ regulator actually changes the nature of our solution. Once regularized, it will represent the tachyon condensation vacuum solution and $\e$ will turn out to be a gauge parameter.

\subsection{Application to the lump solution}

We proceed now to regularize our lump solution.
At every step of our equations in section 2.2 of \cite{BMT} or in the introduction of this paper, we replace $\phi$ with $\phi+\epsilon$, where $\epsilon$ is a small number we will take eventually to 0,
and use the above Schwinger representations wherever we find inverted vectors.

Our lump solution becomes
\be
\psi_{\phi}= c(\phi+\epsilon) - \frac 1{K+\phi+\epsilon} (\phi+\epsilon -\delta \phi) Bc\partial c.
\label{psiphieps}
\ee
This is certainly a solution to the equation of motion since it is simply obtained by replacing
$\phi$ with $\phi+\epsilon$ and it is certainly regular. Moreover
\be
U_{\phi} = 1- \frac 1{K+\phi+\epsilon} (\phi+\epsilon) Bc,
\quad\quad U_{\phi}^{-1} = 1+ \frac 1 K (\phi+\epsilon) Bc.\label{Uphi}
\ee
Therefore $U_{\phi}^{-1}$ remains singular, and the solution is non-gauge for any $\epsilon$.

In proving that (\ref{psiphieps}) is a solution we need
\be
\frac 1{K+\phi+\epsilon}(K+\phi+\epsilon)=1.\label{inverse}
\ee
This is certainly correct for any $\epsilon\neq 0$. Therefore we can assume, by continuity, that it is true also for $\epsilon=0$. This can be confirmed in a weak sense
as follows
\be
&&Tr \left( \frac 1{K+\phi+\epsilon}(K+\phi+\epsilon)\right) =
\int_0^{\infty} dt \,Tr\left( e^{-t (K+\phi+\epsilon)} (K+\phi+\epsilon)\right)=\0\\
&&- \int_0^{\infty}dt\,\frac {\partial}{\partial t} Tr \left(e^{-t (K+\phi+\epsilon)}\right)=
g(0)- Tr \left(e^{-t (K+\phi)}\right) e^{-t\epsilon }\Bigr{|}_{t=\infty}\0\\
&&= g(0)- \lim_{t\to \infty} g(t u) e^{-t\epsilon }.\label{verif1}
\ee
The second term
in the RHS of (\ref{verif1}) vanishes as long as $\Re{\epsilon}>0$ (as we shall always assume).
$g(0)$ is the expected response corresponding to 1 in the RHS of (\ref{inverse}). In fact
\be
g(0)=\lim_{u\to 0} \frac 1{2\sqrt{\pi u}}= \delta(0) = \frac V{2\pi}\label{g(0)}
\ee
The importance of this result should however not be overstimated because of the subtraction necessary
in order to obtain a finite result on the RHS, see in this regard Appendix D.

{\bf Remark 1}. In the above integral (\ref{verif1}) we are not allowed to exchange the $\epsilon \to 0$ limit with integration, because the function $g(tu)$ {\it is not integrable for large $t$}.
\vskip 0.5cm

Let us see next the other conditions mentioned in the introduction. In the solution we find the expression
\be
\frac 1{K+\phi_u+\epsilon} (\phi_u+\epsilon -\delta \phi_u).\nonumber
\ee
We have to check that it is regular. Again it is certainly well--defined for any $\epsilon$ with
$\Im \epsilon \neq 0$. Using a Schwinger representation we choose $\Re{\epsilon}>0$.
The one-point correlator is
\be
&&\langle \frac 1{K+\phi_u+\epsilon} (\phi_u+\epsilon -\delta \phi_u)\rangle=
\langle  \frac 1{K+\phi_u+\epsilon} (\epsilon + u \partial_u \phi_u)\0\\
&&=\epsilon \int_0^\infty dt \, e^{-\epsilon t} \langle e^{-t(K+\phi_u)}\rangle
- \int_0^\infty dt \, e^{-\epsilon t} \frac ut \partial_u \langle  e^{-t(K+\phi_u)}\rangle\0\\
&&= \epsilon \int_0^\infty \frac {dx}u \, g(x)  e^{-\epsilon \frac xu}- \int_0^\infty dx\, \partial_x g(x) \, e^{-\epsilon \frac xu}\0\\
&& =- \int_0^\infty dx\, \partial_x\left( g(x)  e^{-\epsilon \frac xu}\right) =
g(0)-\lim_{x\to\infty} g(x)  e^{-\epsilon \frac xu},\label{verif2}
\ee
where $x=tu$.
As long as $\epsilon,u$ are kept finite, the above limit vanishes and we get
\be
\lim_{\epsilon \to 0}\langle  \frac 1{K+\phi_u+\epsilon} (\phi_u+\epsilon -\delta \phi_u)\rangle=
g(0).\label{verif4}
\ee
If we take the limit $\epsilon\to 0$ first, we get instead
\be
\langle  \frac 1{K+\phi_u} (\phi_u-\delta \phi_u)\rangle=
g(0)-g(\infty).\label{verif5}
\ee

{\bf Remark 2.} This is another example in which we cannot exchange integration with $\e\to 0$ limit,
the reason being the usual one: $g(x)$ is not integrable for large $x$. Therefore the correct regularized result is given by (\ref{verif4}). Such discontinuity of the $\e\to 0$ limit will play a fundamental role in the sequel.
\vskip 0.5cm

Let us consider next $\langle \frac 1{K+\phi_u+\epsilon}\rangle$ which is expected to be singular.
We have
\be
\langle \frac 1{K+\phi_u+\epsilon}\rangle&=& \int_0^\infty dt \, e^{-\epsilon t}
 \langle e^{-t(K+\phi_u)} \rangle=  \int_0^\infty\frac {dx}u g(x)  e^{-\epsilon \frac xu}.\label{verif6}
\ee
The crucial region is at $x\to \infty$. Since $g(\infty)= finite$ the behaviour of this integral
is qualitatively similar to
\be
\sim \frac 1{\epsilon}  e^{-\epsilon \frac xu}\Bigr{|}_M^{\infty}\sim \frac {e^{-\frac{\e M}u}}{\e}\label{verif7}
\ee
for $M$ a large number. The inverse of $\epsilon$ present in this expression makes the integral (\ref{verif6}) divergent, as it is easy to verify also numerically.
This tells us that homotopy operator corresponding to the regularized solution (see below) is well-defined, while if we set $\e=0$ it becomes singular.

As the above examples show, the $\e\to 0$ limit, in general, is not continuous. This is true in particular
for the energy, as we shall see in the next section.

\subsection{Other regularizations}

The regularization we have considered so far in this paper (named $\epsilon$-regularization) is far from unique.
It consists in adding to $\phi_u$ the operator $\e I$. However we are free to add suitable perturbing operators instead of the identity operator $I$ and generate families of solutions. In particular we will consider in the following replacing $\phi_u(s)$ with $\phi_u(s) +\e f(s)I$, where $f(s)$ is some function of $s$. It is easy to prove that these are all solutions to the equation of motion. Some of them are particularly important and simple to deal with, they are defined by the choices
\be 
f_1(s)&=& \theta(s-M) \label{tM}\\
f_2(s)&=& \theta(M-s) \label{Mt}
\ee
where $M$ is some finite number. The first choice gives rise to a regularization that dumps the IR, just as the $\e$ regularization does. Therefore such a family of solutions (depending on $M$ and $\e$) will be gauge equivalent to the tachyon vacuum solution.
The second choice does not affect the IR and gives rise to a family of solutions which are gauge equivalent to the lump. These different regularizations will be discussed in a separate paper, \cite{BGT3}.

\section{Regulated energy}

In this section we calculate the energy of the regularized solution. Our aim is to study
the $\e\to 0$ limit and verify whether it is continuous or not.

The regulated solution is
\begin{align}
\psi_u^{\epsilon}=\lim_{\epsilon\to
0}\Big(c(\phi_u+\epsilon)-\frac1{K+\phi_u+\epsilon}(\phi_u+\epsilon-\delta\phi_u)Bc\partial
c\Big).\label{psiu}
\end{align}

 The energy is proportional to
 \begin{align} \langle \psi_u
\psi_u\psi_u\rangle & = -\lim_{\epsilon\to
0}\Big\langle\Big( \frac
1{K+{\phi}_u+\epsilon} ({\phi}_u+\epsilon+2u) BcKc\Big)^3
  \Big\rangle\label{energy1'}\\
 &=-\lim_{\epsilon\to
0}\int_{0}^\infty dt_1dt_2dt_3{\cal E}_0(t_1,t_2,t_3)e^{-\epsilon
T}\Big\langle ({\phi}_u(t_1+t_2)+\epsilon+2u)\0\\
&\quad\quad\quad\quad\times({\phi}_u(t_1)+\epsilon+2u)({\phi}_u(0)+\epsilon+2u)e^{-\int_0^Tds{\phi}_u(s)}
 \Big\rangle_{C_T},\0
\end{align}
 where $T=t_1+t_2+t_3$. We map the matter parts to the unit disc:
 \begin{align}
  \langle \psi_u\psi_u\psi_u\rangle &=-\lim_{\epsilon\to
0}\int_{0}^\infty dt_1dt_2dt_3{\cal
E}_0(t_1,t_2,t_3)e^{-2uT\big(\ln(\frac {uT}{2\pi})+A+\frac\epsilon
{2u}\big)}\label{energy2}\\
&\quad\quad\times u^3\Big\langle
\Big(X^2(\theta_{t_1+t_2})+2\big(\ln(\frac{uT}{2\pi})+A+1+\frac\epsilon
{2u}\big)\Big)\0\\
 &\quad\quad\times \Big(X^2(\theta_{t_1})+2\big(\ln(\frac{uT}{2\pi})+A+1+\frac\epsilon
{2u}\big)\Big)\0\\
 &\quad\quad\times\Big(X^2(0)+2\big(\ln(\frac{uT}{2\pi})+A+1+\frac\epsilon
{2u}\big)
 \Big)e^{-\int_0^{2\pi}d\theta
 \frac{uT}{2\pi}X^2(\theta)}\Big\rangle_{Disk}.\0
  \end{align}
 Using Appendix D of I and setting $A=\gamma-1+\ln4\pi$, we obtain
 \be
&&\langle \psi_u\psi_u\psi_u\rangle
=-\lim_{\epsilon\to0}\int_{0}^\infty
dt_1dt_2dt_3{\cal E}_0(t_1,t_2,t_3)u^3e^{-\epsilon T}g(uT)\label{energy3}\\
&&\cdot\Big\{8\Big(\frac{h_{2uT}}2+\ln(2uT)+\gamma+\frac\epsilon
{2u}\Big)^3+8G_{2uT}(\frac{2\pi t_1}T)G_{2uT}(\frac{2\pi
(t_1+t_2)}T)G_{2uT}(\frac{2\pi t_2}T)\0\\
&&+4\Big(\frac{h_{2uT}}2+\ln(2uT)+\gamma+\frac\epsilon
{2u}\Big)\Big(G_{2uT}^2(\frac{2\pi t_1}T)+G_{2uT}^2(\frac{2\pi
(t_1+t_2)}T)+G_{2uT}^2(\frac{2\pi
t_2}T)\Big)\Big\}.\0
 \ee
This can also be written as
\begin{align}
 \langle \psi_u \psi_u\psi_u\rangle &=-\lim_{\epsilon\to0}\int_{0}^\infty
dt_1dt_2dt_3{\cal E}_0(t_1,t_2,t_3)u^3e^{-\epsilon
T}g(uT)\Bigg\{\Big(\frac\epsilon
{u}-\frac{\partial_{uT}g(uT)}{g(uT)}\Big)^3\label{energy4}\\
&+2\Big(\frac\epsilon
{u}-\frac{\partial_{uT}g(uT)}{g(uT)}\Big)\Big(G_{2uT}^2(\frac{2\pi
t_1}T)+G_{2uT}^2(\frac{2\pi (t_1+t_2)}T)+G_{2uT}^2(\frac{2\pi
t_2}T)\Big)\0\\
&+8G_{2uT}(\frac{2\pi t_1}T)G_{2uT}(\frac{2\pi
(t_1+t_2)}T)G_{2uT}(\frac{2\pi t_2}T)\Bigg\}.\0
\end{align}

Let us make again a change of variables $(t_1,t_2,t_3)\to (T,x,y)$, where
\be
x&=\frac{t_1}{T},\quad\quad y=\frac{t_2}{T}.\0
\ee
  Then the matter part of the energy can be written as
$$u^3 e^{-\epsilon T}F_\epsilon(uT,x,y),$$
where
\begin{align}
  F_\epsilon(uT,x,y)&=g(uT)\Bigg\{\Big(\frac\epsilon
{u}-\frac{\partial_{uT}g(uT)}{g(uT)}\Big)^3\label{Fepsilon}\\
&  +8G_{2uT}(2\pi x)G_{2uT}(2\pi(x+y))G_{2uT}(2\pi y)\0\\
+& 2 \Big(\frac\epsilon
{u}-\frac{\partial_{uT}g(uT)}{g(uT)}\Big)\Big(G_{2uT}^2(2\pi
x)+G_{2uT}^2(2\pi(x+y))+G_{2uT}^2(2\pi y)\Big) \Bigg\}.\0
\end{align}
The ghost correlator has been given in the introduction.
Making an additional change of coordinate $s=2uT, x\rightarrow y\rightarrow 1-y$, yields finally
\begin{align}
E_0[\psi_u]=\frac 16\lim_{\epsilon\to0}\int_0^\infty ds\;
s^2\int_0^1 dy\int_0^{y} dx\,e^{-\frac{\epsilon s}{2u}}{\cal
E}(1-y,x)\, F_\epsilon(s/2,1-y,x)\label{final1}
\end{align}
with
\begin{align}
  F_\epsilon(s/2,1-y,x)&= \g(s)\Bigg\{\Big(\frac\epsilon
{2u}-\frac{\partial_{s}\g(s)}{\g(s)}\Big)^3\label{Fs}\\
&  +G_{s}(2\pi x)G_{s}(2\pi(x-y))G_{s}(2\pi y)\0\\
+&\frac 12\Big(\frac\epsilon
{2u}-\frac{\partial_{s}\g(s)}{\g(s)}\Big) \Big(G_{s}^2(2\pi
x)+G_{s}^2(2\pi(x-y))+G_{s}^2(2\pi y)\Big) \Bigg\}.\0
\end{align}
where we have set $\g(s)\equiv g(s/2)$.

 \subsection{The energy in the limit $\epsilon\to 0$}

Our purpose here is to study the energy functional $E_{\epsilon}[\psi_u^{\epsilon}]$ . We notice that (\ref{final1},\ref{Fs}) for generic $\epsilon$ can be obtained directly from (\ref{Efinal}) with the following exchanges: $U\to s$, $g(uT)\to e^{-\frac{\epsilon s}{2u}} \g(s)\equiv\tilde \g_{\epsilon}(s,u)$. Summarizing, we can write
\be
E_{\epsilon}[\psi_u^{\epsilon}]&=&\frac 16\lim_{\epsilon\to0}\int_0^\infty ds\;
s^2\int_0^1 dy\int_0^{y} dx\, {\cal
E}(x,y)\,\tilde \g_{\epsilon}(s,u)\Bigg\{\Big(-\frac{\partial_{s}\tilde \g_{\epsilon}(s,u)}{\tilde \g_{\epsilon}(s,u)}\Big)^3\0\\
&&  +G_{s}(2\pi x)G_{s}(2\pi(x-y))G_{s}(2\pi y)\0\\
&+&\frac 12\Big( -\frac{\partial_{s}\tilde \g_{\epsilon}(s,u)}{\tilde \g_{\epsilon}(s,u)}\Big) \Big(G_{s}^2(2\pi
x)+G_{s}^2(2\pi(x-y))+G_{s}^2(2\pi y)\Big) \Bigg\}.\label{final2}
\ee
We are of course interested in the limit
\be
E_0[\psi_u]=\lim_{\epsilon\to0}E_{\epsilon}[\psi_u^{\epsilon}]\label{finallimit}
\ee
but we will see that in fact the energy functional $E_{\epsilon}[\psi_u^{\epsilon}]$ does not depend on $\e$.

The dependence on $\epsilon$ is continuous in the integrand, therefore
a discontinuity in the limit $\epsilon\to 0$ may come only from divergent integrals that multiply $\epsilon$
factors. Now, looking at (\ref{Fs}), we see that we have two types of terms. The first type is nothing but
(\ref{Efinal}), with the only difference that the integrand of $d(2uT)$ is multiplied by
 $e^{-\frac{\epsilon s}{2u}}$.
In the previous sections we have shown that, setting formally $\epsilon=0$ everywhere in (\ref{final1})
and (\ref{Fs}), or in (\ref{final2}), and subtracting the UV singularity, we get a finite integral, i.e. in particular the integrand has integrable behaviour for $s\to\infty$. Therefore this first type of term
is certainly continuous in the limit $\e\to 0$. However with the second type of terms the story is different.
The latter are the terms linear, quadratic or cubic in $\frac\epsilon{u}$ in (\ref{Fs}) (for convenience we will call them $\e$-terms). The factors that
multiply such terms in the integrand may be more singular than the ones considered in the previous section. They may give rise to divergent integrals, were it not for the overall factor $e^{-\frac{\epsilon s}{2u}}$. In the $\epsilon\to 0$ limit these terms generate a (finite) discontinuity through a mechanism we shall explain in due course.

To proceed to a detailed proof we will split the $s$ integration into three intervals:
$0-m$, $m-M$ and $M-\infty$, where $m$ and $M$ are finite numbers, small ($m$) and large ($M$) enough for our purposes. It is obvious that, since possible singularities of the $s$-integral may arise only at $s=0$ or $s=\infty$, the integral between $m$ and $M$ is
well defined and continuously dependent on $\e$, so for this part we can take the limit
$\e\to 0$ either before or after integration, obtaining the same result.

In the sequel we will consider the effect of the  $\e\to 0$ in the UV and in the IR, the only two regions
where a singularity of the mentioned type can arise.

\subsubsection{The $\e$-terms and the $\e\to 0$ limit in the UV}

Here we wish to check that the $\e$-terms do not affect the singularity in the UV, so that the
subtraction in section 4 remains unaltered. It is enough to limit ourselves to the integral in the interval $(0,m)$, where $m$ is a small enough number.
Let us start from the term proportional to
$\g(s)\left(\frac {\e}{2u}\right)^3 e^{-\frac{\epsilon s}{2u}}$, coming from
the first line of (\ref{Fs}). To simplify the notation we will denote $\frac{\epsilon }{2u}$ simply by $\eta$.
Since, near 0, $\g(s)\approx 1/\sqrt{s}$, this first term gives rise to the integral (for $s\approx 0$),
\be
&&\int_0^m ds s^{\frac 32} \eta^3e^{-\eta s} \label{zero1}\\
&& \sim\left(-\eta\, e^{-\eta s} \frac{\sqrt{s} (3+ 2 s \eta)}{2} +
\sqrt{\eta} \frac {3 \sqrt{\pi} Erf\left(\sqrt{{\eta s}}\right)}{4 } \right)\Bigr{|}_0^m.\0
\ee
Since the error function $Erf(x)\approx x$ for small $x$, it is evident that this expression vanishes both at
$s=0$ and in the limit $\e\to 0$.

The next term to be considered is $  {\eta}^2\partial_s \g(s)$, which leads to the integral
\be
&&\int_0^m ds s^{\frac 12}  {\eta}^2e^{-\eta s}
\sim\left(-\eta\, e^{-\eta  s}\sqrt{s} +
\sqrt{\eta} \frac {\sqrt{\pi} Erf\left( {\eta s}\right)}{2} \right)\Bigr{|}_0^m\ \label{zero2}
\ee
which again vanishes in the $\e\to 0$ limit.

The following term leads to the integral
\be
&&\int_0^m ds \frac 1{\sqrt{ s}}\,\eta\,e^{-\eta s}
\sim\left( \sqrt{\pi\eta}   Erf\left(\sqrt{\eta s}\right) \right)\Bigr{|}_0^m \label{zero3}
\ee
which vanishes as well in the $\e\to 0$ limit.

Finally the term linear in $\e$ coming from the last two lines of (\ref{Fs}) gives rise to an UV
behaviour $\sim  s^{-\frac 12}    \eta$. Therefore the relevant UV integral is
similar to (\ref{zero3}) and we come to the same conclusion as above.

In conclusion the $\e$-terms do not affect the UV behaviour of the energy integral, and in the $\e\to 0$ limit they yield evanescent contributions.

Finally let us consider what remains after discarding the $\e$-terms. From section 4 the behaviour for $\e\approx 0$ is the following
\be
\int_0^m ds \, \frac {e^{-\eta s}}{s^{\frac 32}}=\left(-2 \frac{e^{-\eta s}}{\sqrt s}  -2
\sqrt{\pi\eta} Erf\left(\sqrt{ \eta s}\right) \right)\Bigr{|}_0^m.\label{zero4}
\ee
In the limit $\e\to 0$ the second term vanishes. The first term gives the expected UV singularity we have subtracted away in section 4.

\subsubsection{The $\e$-terms and the $\e\to 0$ limit in the IR}

There is a chance, with $\e$-terms, that the corresponding integrals diverge or produce
negative powers of $\e$, leading to finite or divergent contributions in the limit $\e\to 0$.

Let us start again from the term proportional to
$\g(s)\eta^3 e^{-\eta s}$, coming from
the first line of (\ref{Fs}). The integration in $x,y$ gives a finite number. $\g(s)$ tends to a constant for $s\to\infty$.  To appreciate qualitatively the problem we replace $\g(s)$
by a constant and integrate between $M$ and infinity, $M$ is chosen large enough so that
$\g(s)= const$ is a good approximation. The integral is proportional to
\be
&&\int^{\infty}_M ds \,s^2 \eta^3e^{- \eta s}\sim -  e^{-\eta s}\left(2 + 2\eta s+ \eta^2 s^2\right)\Bigr{|}_M^{\infty}
 \label{inf1}\\
 &&= e^{-\eta M}\left(2+2 {\eta M} + \eta^2M^2\right)\0
\ee
which does not vanish in the limit $\e\to 0$.

Let us notice that, if we consider an additional term in the asymptotic expression for
$\g(s)$, say $\g(s)= a + \frac bs+\ldots$, the additional $\frac 1s$ term contributes
to the RHS of (\ref{inf1}) an additional term $\sim \eta\, e^{- \eta M}
(1 +\eta M)$, which vanishes in the $\e\to 0$ limit. The more so for the
next approximants. This is always the case in the following discussion, therefore  considering the asymptotically dominant term will be enough for our purposes.

Let us consider next the term proportional to $\eta^2\partial_s \g(s)$.
For large $s$  we have $\partial_s \g(s) \sim \frac 1{s^2}$. Therefore the integral to be considered is
\be
&&\int^{\infty}_M ds \eta^2e^{-\eta s}\sim -  e^{-\eta s} \eta \Bigr{|}_M^{\infty}
= e^{-\frac{\epsilon M}u}\frac {\e}u \label{inf2}
\ee
which vanishes in the limit $\e\to 0$.

The linear term in $\e$ coming from the first line of (\ref{Fs}), that is the term proportional to
$\eta \frac{(\partial_s \g(s))^2} {\g(s)}$, leads to a contribution that can be qualitatively represented by the integral
\be
&&\int^{\infty}_M ds \,\frac 1{s^2} \eta e^{- \eta s}\sim -  \eta \left(\frac{e^{-\eta s}}s - \eta {Ei\left(-\eta   s \right)} \right)  \Bigr{|}_M^{\infty}
 \label{inf3}\\
 &&=\eta e^{-\eta M}- \eta^2 {Ei\left(- \eta M\right)}\0
\ee
which vanishes in the limit $\e\to 0$, because the exponential integral function $Ei(-x)$ behaves like $\log x$ for small $x$.

Finally let us consider the term linear in $\e$ coming from the last two lines of (\ref{Fs}), i.e.
\be
\g(s) \eta\, \Big(G_{s}^2(2\pi
x)+G_{s}^2(2\pi(x-y))+G_{s}^2(2\pi y)\Big).\label{inf4}
\ee
The integration of $x,y$ of the $G^2_s$ terms in brackets gives a contribution behaving at infinity as $1/s^2$ (see section 5). Therefore the relevant $s$ contribution for large $s$ is
 \be
\int^{\infty}_M ds  \eta e^{- \eta s}\sim -  e^{-\eta s}  \Bigr{|}_M^{\infty}= e^{-{\eta M}}\0
\ee
which is nonvanishing in the limit $\e \to 0$.

Therefore we have found two nontrivial $\e$--terms, the first and the last ones above. Let us call them
$\alpha$ and $\beta$, respectively. They  do not vanish in the limit $\e\to 0$, thus they may survive this limit and represent a finite difference between taking $\e\to 0$ before and after the
$s$-integration. It is therefore of utmost importance to see whether the overall contributions of these two terms survives. This turns out to be the case.

From section 5 one can check that the precise form of the first term in question is
\be
-\frac 1{4\pi^2} \lim_{\e\to 0} \int_0^{\infty} ds\,s^2 \g(s) \eta^3 e^{-\eta s}.\label{1st}
\ee
If one knows the asymptotic expansion of the integrand for large $s$, it is very easy to extract the exact $\e\to 0$ result of the integral. The
asymptotic expansion of $\g(s)$ is $\g(s)\approx    1+\frac 1{24 s} + \frac 1{1152 s^2}+\ldots $. Integrating term by term from $M$ to $\infty$, the dominant one gives
\be
-\frac 1{4\pi^2}e^{-\eta M } (2 +2 M \eta+M^2 \eta^2) \label{0term1}
\ee
which, in the  $\e\to 0$ limit, yields $-\frac 1{2\pi^2}$. The next term gives $\sim e^{- M \eta} ( \eta (1+M \eta)$, which vanishes in the  $\e\to 0$ limit, and so on. So the net result of
the integral (\ref{1st}) in the  $\e\to 0$ limit is $-\alpha$, where
\be\label{alpha}
\alpha\equiv \frac 1{2\pi^2}.
\ee

For the $\beta$ term (the one corresponding to (\ref{inf4})) we have
\be
-\beta=\frac 1{12} \lim_{\e\to 0} \int_0^{\infty}ds\, s^2 \g(s) \eta e^{-\eta s}
\left( -\frac a{s^2}+\cdots\right), \label{2nd}
\ee
where ellipses denote terms that contribute vanishing contributions in the $\e\to 0$ limit
and $a$ is the (overall) coefficient of the inverse quadratic term in (\ref{GU21}).
The problem is to compute the latter. With reference to the enumeration in section 6.2, the term 1) has the asymptotic expansion
\be
\sim e^{-\eta s}\left(-\frac{3 \eta}{8 \pi ^2}-\frac{\eta}{64 s\pi ^2}- \cdots\right).\label{2term1}
\ee
Integrating from $M$ to $\infty$ and taking the $\e\to 0$ limit, this gives $-\frac{3}{8 \pi ^2} $.
Proceeding in the same way,  term 2) of section 6.2 gives $\frac{3 }{4\pi ^2} $ and term 3) yields
$\frac{1}{4 \pi ^2} $. So altogether we have $\frac 5{8\pi^2}$ for the three terms contributing to
(\ref{2nd}) considered so far.

It remains term 4) of section 6.2. This corresponds to
the contribution of $RK(p,U)-\frac 1{8\pi(p^2-1)}$.  One must explicitly sum over $p$ in order to know the asymptotic expansion in $U$. This has not been possible so far analytically. However Mathematica can compute
the coefficient of $1/U^2$ in the asymptotic expansion for large $U$ to a remarkable accuracy. The coefficient
turns out to be -0.064317, with an uncertainty only at the fifth digit. Within the same uncertainty this corresponds to the analytic
value $- \frac 1{4 \pi} \left(\gamma + 1/3 \log 2\right)$. We therefore set
\be \label{beta}
\beta= -\frac 5{8\pi^2}+\frac 1{\pi^2}\left(\gamma +\frac 13 \log 2\right).
\ee

So the overall contribution of the $\e$-terms in the $\e\to 0$ limit is
\be
-\alpha-\beta=-\frac 1{2\pi^2}+ \frac 5{8\pi^2}-
\frac 1{\pi^2}\left(\gamma +\frac 13 \log 2\right)\approx -0.0692292,\label{fregati}
\ee
which is accurate up to the fourth digit.

Let us consider now what remains apart from the $\e$-terms. The integrand takes the form
\be
\int_0^{\infty} ds\, F(s) \,e^{-\eta s},\label{whatremains}
\ee
where $F(s)$ represents the integrand when $\e=0$, i.e. the total integrand analyzed in section 5. We have already argued that the integration over $s$ and the limit $\e\to 0$ can be safely exchanged, which yields
the already found value of 0.0693926.

Concluding we have
\be
E^{(s)}_0[\psi_u]=\lim_{\epsilon\to0}E^{(s)}_{\epsilon}[\psi_u^{\epsilon}]\approx 0.000163,\label{ultimatelimit}
\ee
where again the superscript $^{(s)}$ means that the UV singularity has been subtracted away.

This teaches us two lessons. First, that the regularized solution is the tachyon vacuum solution. This is true not only in the limit $\e\to 0$ but also for nonvanishing $\e$  and, consequently, $\e$ plays the role of a gauge parameter. This conclusion can be reached only via numerics, for the calculation with $\e\neq 0$ cannot be analytical, and thus the result is less precise. But it is nevertheless significant to see the values $E^{(s)}_{\epsilon}[\psi_u^{\epsilon}]$
for various values of $\eta$ in Table 3.
\TABLE[ht]{
$\begin{matrix}\quad &\eta: & 2\,&1\,&0.1 \, &  0.01\,& 0.001\\
\quad&E^{(s)}_{\epsilon}[\psi_u^{\epsilon}]:\quad&4\times 10^{-6}\quad&  8\times 10^{-6} \quad&0.000603 \quad& 0.001832\quad& 0.007360
\end{matrix}$
\caption{Samples of $E^{(s)}_{\epsilon}[\psi_u^{\epsilon}]$ }
}
These values are close to 0, but with an accuracy that worsens for decreasing $\eta$.
It is worth spending a few words on the numerical origin of this fact. For instance the quadratic $\e$--terms
are characterized by an integrand consisting of two factors: the first is a sort of Gaussian, whose
maximum increases in value and position like the inverse of $\eta$; on the contrary the other factor decreases,
with the overall result that the integral in $s$ varies slightly with $\eta$. This explains why
the $\e$--terms, which are negative, kill completely the overall positive contribution coming from the other terms. The trouble with this scheme is that for smaller $\eta$'s the integral must be evaluated over larger and larger intervals of $s$ in order to approximate its true value, and this clashes inevitably with the
computing capacities of Mathematica. This explains the worsening performance for decreasing $\eta$.

The second lesson we learn is that, since at this point we can assume the true value of $E^{(s)}_\e[\psi_u]$ to be 0, and since the value (\ref{fregati}) is much more accurate than (\ref{finalresult}), we can take for the latter the more reliable value
\be
E^{(s)}[\psi_u]\approx  0.0692292\label{ultimateresult}
\ee
which we can consider at this point to be exact (even though this will not play any role in the determination of the lump energy). It differs from (\ref{finalresult}) by 2 per mil. Therefore, after all, our numerical evaluation in section 6 was not so bad.
Stated differently, the whole procedure of this section is nothing but a more reliable way to compute the
energy functional (\ref{Efinal}). 

We have already remarked that (\ref{ultimateresult}) differs from the theoretical value
(\ref{TD24}) of the lump energy by 27\%. This is not the expected lump energy. But now we have everything at hand
to explain the puzzle.

\section{The lump and its energy}

In the previous sections we have found various solutions to the equation of motion $Q\psi+\psi\psi=0$ at the perturbative vacuum.
One is $\psi_u$ with UV--subtracted energy (\ref{ultimateresult}), the others are the $\psi_\e$'s with generic $\e$ and vanishing UV--subtracted energy. Using these we can construct a solution to the EOM  {\it at the tachyon condensation vacuum}.

The equation of motion at the tachyon vacuum is
\begin{align}\label{EOMTV}
{\cal Q}\Phi+\Phi\Phi=0,\quad {\rm where}~~{\cal
Q}\Phi=Q\Phi+\psi_\epsilon\Phi+\Phi\psi_\epsilon.
\end{align}
We can easily show that
\begin{align}\label{psiupsie}
\Phi_0=\psi_u-\psi_\epsilon
\end{align}
is a solution to (\ref{EOMTV}). The action at the tachyon vacuum is
\begin{align}\label{TVaction}
-\frac12\langle{\cal
Q}\Phi,\Phi\rangle-\frac13\langle\Phi,\Phi\Phi\rangle.
\end{align}
Thus the energy is
\begin{align}\label{TVlumpenergy}
E[\Phi_0]=-\frac16\langle\Phi_0,\Phi_0\Phi_0\rangle=
-\frac16\big[\langle\psi_u,\psi_u\psi_u\rangle-\langle\psi_\epsilon,\psi_\epsilon\psi_\epsilon\rangle
-3\langle\psi_\epsilon,\psi_u\psi_u\rangle+3\langle\psi_u,\psi_\epsilon\psi_\epsilon\rangle\big].
\end{align}
Eq.(\ref{psiupsie}) is the lump solution at the tachyon vacuum, therefore, this energy must be the energy
of the lump.

We have already shown that $-\frac16 \langle\psi_u,\psi_u\psi_u\rangle^{(s)}=\alpha+\beta$ and that $\langle\psi_\epsilon,\psi_\epsilon\psi_\epsilon\rangle^{(s)}=0$, after subtracting the UV singularity.
It remains for us to compute the two remaining terms, which we will do in the next subsection. But, before,
let us remark one important aspect of (\ref{TVlumpenergy}). The UV subtractions are the same in all terms, therefore they neatly cancel out.

\subsection{Two more terms}

The two terms $\langle\psi_\epsilon,\psi_u\psi_u\rangle$ and $\langle\psi_u,\psi_\epsilon\psi_\epsilon\rangle$
can be calculated in the same way as the other two, and we limit ourselves to writing down the final result:
\be
\langle\psi_\epsilon,\psi_u\psi_u\rangle&=&- \int_0^\infty ds\;
s^2\int_0^1 dy\int_0^{y} dx\,e^{-\eta s}{\cal
E}(1-y,x) \,e^{\eta sy}\, \g(s)\label{Fs2}\\
&&\cdot\Bigg\{\Big(\eta -\frac{\partial_{s}\g(s)}{\g(s)}\Big)\Big(-\frac{\partial_{s}\g(s)}{\g(s)}\Big)^2
 +G_{s}(2\pi x)G_{s}(2\pi(x-y))G_{s}(2\pi y)\0\\
&&\,+\frac 12\Big(\eta-\frac{\partial_{s}\g(s)}{\g(s)}\Big)G_{s}^2(2\pi x)+ \frac 12\Big(-\frac{\partial_{s}\g(s)}{\g(s)}\Big)\Big(G_{s}^2(2\pi
y)+G_{s}^2(2\pi (x-y))\Big) \Bigg\}.\0
\ee
and
\be
\langle\psi_u,\psi_\e\psi_\e\rangle&=&- \int_0^\infty ds\;
s^2\int_0^1 dy\int_0^{y} dx\,e^{-\eta s}{\cal
E}(1-y,x) \,e^{\eta sx}\, \g(s)\label{Fs3}\\
&&\cdot\Bigg\{\Big(\eta -\frac{\partial_{s}\g(s)}{\g(s)}\Big)^2\Big(-\frac{\partial_{s}\g(s)}{\g(s)}\Big)
 +G_{s}(2\pi x)G_{s}(2\pi(x-y))G_{s}(2\pi y)\0\\
&&\,+\frac 12\Big(-\frac{\partial_{s}\g(s)}{\g(s)}\Big)G_{s}^2(2\pi(x-y))+ \frac 12\Big(\eta-\frac{\partial_{s}\g(s)}{\g(s)}\Big)\Big(G_{s}^2(2\pi
x)+G_{s}^2(2\pi y)\Big) \Bigg\}.\0
\ee
Although we believe that the results below holds for any value of $\e$, the calculation for generic $\e$
is beyond our present means, therefore from now on in this section we will condider only the $\e\to 0$ limit.
A bonus of this limit is that it will give us analytic results.
In this limit the factors $e^{\eta sy}$ and $e^{\eta sx}$, present in (\ref{Fs2},\ref{Fs3}, respectively), are irrelevant. In fact the integration over  $y$, without this factor, is finite. Therefore we know from above that the integration over $y$ with
$e^{\eta sy}$ inserted back at its place is continuous in $\e$ for $\e\to 0$ (one can check that the subsequent integration over $s$ does not lead to any complications). Therefore we can ignore these factors in the two integrals above.

The integrals are of the same type as those analyzed in the previous section.
Of course they will have both the contribution that comes from setting $\e=0$, which is proportional to
$\alpha+\beta$, like for $\langle\psi_\e,\psi_\epsilon\psi_\epsilon\rangle$. But there are important differences as far as the $\e$ terms are concerned.
First of all we remark that, in both integrals, the first term in curly brackets does not contain the cubic term in $\e$. Therefore, according to the analysis in the previous section, the $\alpha$ contribution will not be present in either term. On the contrary the $\beta$ contribution, which comes from the last line in both, will. To evaluate it there is no need of new explicit computations. Upon integrating over $x,y$ one can easily realize that the three terms proportional to $G_{s}^2(2\pi x)$, $G_{s}^2(2\pi y)$ and $G_{s}^2(2\pi (x-y))$, give rise to the same contribution. So, when we come to $\e$--terms, each of them will contribute $\frac 13$ of the $\beta$ contribution already calculated in the previous section.
Summarizing: after subtracting the UV singularity, we will have
\be
\frac 16 \langle\psi_\epsilon,\psi_u\psi_u\rangle^{(s)}&=& -\alpha-\beta+\frac 13 \beta,\label{euu}\\
\frac 16 \langle\psi_u,\psi_\e\psi_\e\rangle^{(s)}&=& -\alpha-\beta+\frac 23 \beta.\label{eeu}
\ee

\subsection{Final result}

Let us now collect all the results in (\ref{TVlumpenergy}). The lump energy above the tachyon vacuum is
\be
E[\Phi_0]=\alpha +\beta +0 + 3(-\alpha-\beta+\frac 13 \beta)-3(-\alpha-\beta+\frac 23\beta)=\alpha=
\frac 1{2\pi^2}.\label{lumpenergy}
\ee
This coincides with the expected theoretical value (\ref{TD24}).

As one can see there is no need to know the value of $\beta$, which, anyhow, we have explicitly computed in
the previous section. We stress that the exact value (\ref{lumpenergy}) was computed analytically
in the previous section, see eqs.(\ref{1st},\ref{0term1}), and it is determined by the asymptotics
of $\g(s)$. Moreover we recall another fundamental aspect of (\ref{lumpenergy}): the UV subtractions of the various terms in (\ref{TVlumpenergy}) exactly cancel out.

For completeness it should be added that the result (\ref{lumpenergy}) is based on the assumption we made
in the last section (before eq.(\ref{ultimateresult})) that $E^{(s)}_\e[\psi_u]\equiv 0$.
This was proved in part with numerical methods, but its validity is imposed by consistency.
Finally a few words concerning the $\e$-regularization. It is evident from the above that lump energy comes from the asymptotic region in $s$. This is the region which is precisely suppressed by the $e^{-\eta s}$ factor produced by the $\e$-regularization. It is therefore not surprising that, modulo the UV subtraction, $\psi_\e$
represents a tachyon condensation vacuum solution. 

It is clear that $\e$ plays the role of a gauge parameter,
although the results we have derived are more easily obtained in the $\e\to 0$ limit. There is however no doubt
that the RHS of (\ref{TVlumpenergy}) is independent of the value of $\e$. To show this let us start from the  
following remark. Using the equation of motion, the last two terms of the RHS of (\ref{TVlumpenergy}) could be replaced by $ 3\langle \psi_\e, Q\psi_{ u}\rangle-3\langle\psi_{ u},Q\psi_\e\rangle$.
Formally `integrating by part' (that is moving $Q$ from $\psi_{ u} $ to $\psi_\e$ in the first term) this may seem to give 0. However such conclusion would be incorrect because the correlators in question are UV divergent and an integration by part is not allowed. One would have to regularize them first, but at that point the form (\ref{TVlumpenergy}) is much handier. However this remark leads us to an interesting conclusion. The obstruction to integrating by part is the UV divergence or the corresponding subtraction, which, as we have seen, are $\e$-independent. Therefore the value
of  $ 3\langle \psi_\e, Q\psi_{u}\rangle-3\langle\psi_{u},Q\psi_\e\rangle$ must be $\e$-independent as well. Since $\langle \psi_\e, \psi_\e \psi_\e\rangle$ is also $\e$-independent
as it is expected (and shown numerically in section 8), one is led to conclude that the RHS of (\ref{TVlumpenergy} ) does not depend on $\e$.

\section{Erratum}

After this paper was published in JHEP, JHEP 08(2011)158, a reconsideration of all the problems tackled in it  led us to \cite{BGT3}. In the latter we confirm all the results of this paper as well as of \cite{BGT2}, but we correct the interpretation of $\e$ as a gauge parameter contained in sec. 7.2 and in the last paragraph of 9.2. In \cite{BGT3} we provide evidence that $\langle\psi_\epsilon,\psi_\e\psi_\e\rangle^{(s)}$, $\langle\psi_u,\psi_\e\psi_\e\rangle^{(s)}$ and $\langle\psi_\epsilon,\psi_u\psi_u\rangle^{(s)}$ all depend on $\e$. Consequently $\e$ is simply a regulator and cannot be interpreted as a gauge parameter. The only meaningful results are obtained in the limit $\e\to 0$. Concerning the claim that ``The obstruction to integrating by part is the UV divergence or the corresponding subtraction, which ... are $\e$-independent.'' in the last paragraph of sec.9.2, it is true, but this
does not lead by itself to the implicit conclusion that one can integrate by part the expression $ \langle \psi_\e, Q\psi_{u}\rangle-\langle\psi_{u},Q\psi_\e\rangle$ and get 0, because the UV subtraction is applied to the three-points correlators, not to the string field $\psi_,\psi_\e$, to which $Q$ applies.

\acknowledgments 

A paper by T.Erler and C.Maccaferri, \cite{EM}, will be posted simultaneously to the present paper. It deals with the same problem with a different method and leads to the same results as ours. We would like to thank the authors
for keeping us informed of their progress. The critical remarks of C.~Maccaferri in the course of our
research have been also very helpful.
One of us (L.B.) would like to thank the Yukawa Institute for Theoretical Physics, Kyoto, where a large part of this work has been carried out, for hospitality and financial support, and
D.D.T. would like to thank SISSA for the kind hospitality during part of this research. The work of D.D.T. was supported by the Korean Research Foundation Grant funded by the Korean Government
with grant number KRF 2009-0077423.. 

\vspace{0.5cm}

\section*{Appendix}
\appendix
\section{Integration in the first wedge. Quadratic term.}

We examine here in more detail the integration of the term containing $RK(p,U)$  in the first wedge, as anticipated in sec. 6.1. The first wedge in the $(p,U)$ plane is delimited by the $U$ axis and by the ray $(\epsilon U,U)$, where $\epsilon$ is a finite small number.
In this wedge the expansion of $RK(p,U)$ is given by:
\be
RK'(p,U)&=&  -\frac{1}{16 \pi } \frac {\ln U}{U^3}
+\frac{-7+12 p+7 p^2+6 \left(p^2-1\right) \psi\left(\frac{1+p}{2}\right)}{96 \left(p^2-1\right) \pi }\frac 1{U^3}\0\\
&&- \frac{p \left(-1+2 p+p^2\right)}{16 \pi\left(p^2-1\right)}
\frac 1{U^4}-\frac{1 + p^2}{32 \pi}\frac {\ln U}{U^5}\0\\
&&+ \frac{-103-40 p^2+240 p^3+143 p^4+60 \left(-1+p^4\right) \psi\left(\frac{1+p}{2}\right)}{1920 \pi\left(p^2-1\right) } \frac 1{U^5}\0\\
&&+ \frac{p-2 p^4-p^5}{16 \pi\left(p^2-1\right)}\frac 1{U^6}- \frac{3+10 p^2+3 p^4}{256 \pi }
\frac {\ln U}{U^7}+\cdots,\label{RKpUapp1}
\ee
where we have dropped the first two terms of $RK(p,U)$, because they have been dealt with exactly in section 5.1 (the first when summed over $p$ contributes to cancel the dangerous $\frac 1U$ dependence,
the second can be summed exactly over $p$ leading to a finite coefficient in front of $\frac 1{U^2}$, giving an integrable term for large $U$).
In order to integrate this over $p$ and $U$ we will take the dominant terms (the potentially dangerous ones, as we will see) for large $p$,
\be
RK'(p,U)&\approx& \frac 1{8\pi}\Bigl( - \frac 12 \frac{\ln U}{U^3}+ \frac 12 \ln p \frac 1{U^3} - \frac 12 \frac p{U^4}\label{RKpUapp2}\\
&&  + \frac 14 p^2 \frac {\ln U}{U^5}+\frac 14 p^2\ln p
\frac 1{U^5}-\frac 12 \frac {p^3}{U^6}- \frac 3{32} p^4 \frac {\ln U}{U^7}+\cdots\Bigr).\0
\ee

Integrating in $p$ up to $\epsilon U$ for some small but finite number $\epsilon$ one gets,
\be
\int^{\epsilon U} dp RK'(p,U) &=& \frac 1{16\pi} \Bigl( - \epsilon \frac{\ln U}{U^2}+ \epsilon(\ln(\epsilon U)-1) \frac 1{U^2} - \frac 12 \epsilon^2 \frac 1{U^2} + \frac 1{6}\epsilon^3 \frac {\ln U}{U^2}\label{RKpUapp3}\\
&& +\frac 1{18}
(3\ln(\epsilon U) -1) \frac 1{U^2}- \frac 14 \frac {\epsilon^4}{U^2}-
\frac 3{16} \frac {\epsilon^5}5 \frac {\ln U}{U^2}+\cdots\Bigr).\0
\ee
It is evident that in the RHS we have two numerical series, proportional to $\frac 1{U^2}$
and $\frac {\ln U}{U^2}$, respectively, both strongly convergent because $\epsilon$ can be taken
much smaller than 1. Integrating next over $U$ we get a finite result, because both $\frac 1{U^2}$ and $\frac {\ln U}{U^2}$ are integrable.

\section{Integration in the first wedge. Cubic term.}

We examine here in more detail the integration of the $SK(p,U)$ term
in the first wedge as anticipated in section 5.2. The relevant
expansion for $SK(p,U)$, for asymptotic $p$ as well as $U$, is
(discarding the first term in the RHS of (\ref{SKUasym}) as it is
treated exactly in sec. 5.2) 
\be SK'(p,U)=\frac 1{32 \pi}\left( -
\frac 1{p}\frac 1{U^4} + \frac 1{U^5}-\frac p{U^6}+ \frac
{p^2}{U^7}+\cdots\right).\label{SKpUapp1} \ee Integrating over $p$
up to $\epsilon U$: \be \int^{\epsilon U}dp\, SK'(p,U)&\approx&\frac
1{32 \pi}\left( - \frac {\ln(\epsilon U)}{U^4} + \frac
{\epsilon}{U^4}-\frac {\epsilon^2}{2U^4}+ \frac
{\epsilon^3}{3U^4}+\cdots\right)\label{SKpUapp2} \ee Apart from the
first term in the RHS, which will be discussed in a moment,  we see
that the RHS is a numerical convergent series (because $\epsilon$ is
a finite number which can be chosen to be much smaller than 1)
multiplying the factor $U^{-4}$, i.e. a finite number times
$U^{-4}$. When multiplied by $U^2g(U)$ for large $U$, this produces
an integrable term in $U$.

The first term in the right hand side of (\ref{SKpUapp2}), which
comes in the  p-non-asymptotic form from the second term in the RHS
of (\ref{SKUasym}), is a worrying term, because it increases as
$\epsilon$ becomes small. This term is in connection with a $\frac
{\ln U}{U^4}$ behavior. Such a behavior has been met previously
only in the asymptotic expansion of the $E_0^{(3)}(U)$ term, see
eq.(\ref{approx2}).  $E_0^{(3)}(U)$ contains the terms coming from
the one-cosine and two-cosines angular integrations in the cubic
term. Thus it is natural to search for terms corresponding to the
first term in the RHS of (\ref{SKpUapp2}) among the one-cosine and
two-cosines contributions. This will be rewarding because we will
find an exact cancelation. To see this let us write again (only)
the relevant part of the asymptotic expansion \be \frac {96}{\pi}
SK'(p,U)\sim - \frac 3{\pi^2} \frac 1p \frac 1{U^4}+
\cdots.\label{SKterm1} \ee The corresponding terms coming from the
one-cosine and two-cosines contribution are, respectively, \be {\rm
one-cosine}\sim - \frac 3{\pi^2} \frac 1p \frac 1{U^4}+
\cdots\label{onecos} \ee and
 \be
{\rm two-cosines}\sim + \frac 6{\pi^2} \frac 1p \frac 1{U^4}+
\cdots.\label{twocos} \ee They cancel exactly. On the other hand it
is easy to see that there are no other contributions of the type $
\frac 1p \frac 1{U^4}$ from the remaining three-cosines terms. This
means, on one hand, that we do not have to worry about the first
term in the right hand side of (\ref{SKpUapp2}), on the other hand,
that very likely the term proportional to $\frac {\ln U}{U^4}$ in
(\ref{approx2}), which is anyhow integrable, would not be there if
all the summations could be done analytically down to the end.

\section{D-brane tension and normalization}

In this appendix we will justify the value for the D24 brane tension
in eq.{\ref{TD24}). Let us start from the normalization conventions
used in \cite{Schnabl05,ErlerSchnabl,Okawa1}. They are coherent with
the conventions in Polchinski's book, vol.I, \cite{Polchinski}. The
result for the D25-brane tension (with $\alpha'=1$) \be T_{D25}=
\frac 1{2 \pi^2}\label{T25} \ee was derived by Okawa in
\cite{Okawa02}, appendix A, according to Polchinski's book's
conventions. In the latter the normalization conventions for the
delta function are given in eq.(4.1.15). In the one-dimensional case
they are \be \langle 0,k|0,k'\rangle = 2\pi \delta(k-k')\label{Pol1}
\ee which means in particular 
\be 
\langle 0|0\rangle = 2\pi
\delta(0)= {\EuScript V},\label{Pol2} \ee 
where 
\be \delta(k)= \int
\frac {dx}{2\pi} e^{ipx}.\label{delta} 
\ee

According to these conventions the D24 brane tension must be
\be
{\EuScript T}_{D24}= \frac 1{\pi}.\label{ETD24}
\ee

However in this paper and in I we have been using a different normalization, where
it is understood that
\be
\lim_{u\to 0} \frac 1{2\sqrt{\pi u}}= \delta(0)= \int  \frac {dx'}{2\pi}=\frac {V}{2\pi}
=<0|0>.\label{BSFT}
\ee
where the prime in $dx'$ means that length is measured in different units with respect to (\ref{delta}).
Therefore there is a factor of $2\pi$ between such conventions for the volume, $ V= 2\pi  {\EuScript V}$, and consequently the inverse of this factor for the energy density\footnote{The subtraction of the infinite UV term in section 5 confirms this. In fact, in order to get a finite result, we have subtracted there an overall divergent term that can be written as follows
\be
\frac {15}8 \frac 1{2\pi^2} \frac {V}{2\pi}= \frac {15}8 \frac {V}{4\pi^3}\0
\ee
Apart from the renormalization factor $\frac {15}8$, this tells us that the D25 brane tension is, with our conventions, $\frac 1 {4\pi^3}$ instead of $\frac 1 {2\pi^2}$.}.
It follows that for us the expected result for the D24 brane tension is
\be
T_{D24}= \frac 1{2\pi^2}\label{T'D24}.
\ee

\section{Problems with the equation of motion and the Schwinger representation}

One of the equations used in I, in order to show that our solution
$\psi_\phi$ satisfies the SFT equation of motion, is the following
\begin{align}\label{IDD0}
\frac1{K+\phi}(K+\phi)=I.
\end{align}
Since this equation has been the source of a debate, we would like to use this appendix to explain our point of view in some detail. This problem dealt with in this Appendix refers to \cite{BMT} rather than the present paper, however it is only thanks to the 
results of this paper that some of the issues of \cite{BMT} can be made clear.

With respect to eq.(\ref{IDD0}) a problem arises because, in the case $\phi=\phi_u$, introduced in section 2, we need a Schwinger representation in order to be able to compute correlators. When we use a Schwinger representation, the identity  
\begin{align}\label{IDD}
\frac1{K+\phi_u}(K+\phi_u)=I,
\end{align}
would seem not to be satisfied. To illustrate the problem, let us
calculate the overlap of both the left and the right hand sides of
\eqref{IDD} with $Y=\frac12\partial^2c\partial cc$. The right hand
side is trivial and, in our normalization, it is
\begin{align}\label{LHS}
{\rm Tr}(Y\cdot I)=\lim_{t\to0}\langle Y(t)\rangle_{C_t}\langle
1\rangle_{C_t}=\frac V{2\pi}.
\end{align}
To calculate the left hand side we need the Schwinger representation
\begin{align}\label{RHS}
{\rm Tr}\big[Y\cdot
\frac1{K+\phi_u}(K+\phi_u)\big]=\int_0^{\infty}dt{\rm Tr}\big[Y\cdot
e^{-t(K+\phi_u)}(K+\phi_u)\big]
\end{align}
To evaluate it one is naturally led to make the replacement
\begin{align}\label{naive}
e^{-t(K+\phi_u)}(K+\phi_u)\to -\frac d{dt}e^{-t(K+\phi_u)}
\end{align}
and obtain
\begin{align}\label{wrong}
{\rm Tr}\big[Y\cdot
\frac1{K+\phi_u}(K+\phi_u)\big]=g(0)-g(\infty)=\frac
V{2\pi}-g(\infty),
\end{align}
which is different form \eqref{LHS} because $g(\infty)$ is nonvanishing.
The latter relation is often written in a stronger form
\be 
\int_0^\infty dt\, e^{-t(K+\phi_u)}(K+\phi_u)= 1- \Omega_u^\infty, \quad\quad  \Omega_u^\infty=\lim_{\Lambda\to\infty} e^{-\Lambda(K+\phi_u)}\label{1-Omega}
\ee
This (strong) equality, however, has to be handled with great care. 

Before dealing with the contradiction between \eqref{LHS} and \eqref{wrong}, it is useful to understand, on an independent ground, that eq.(\ref{IDD0}) must be true. Let us start from the observation
that $K+\phi_u$ is a vector in an infinite dimensional space. Defining its inverse by means of a Schwinger representation is one possibility, but not the only one. In fact $K+\phi_u=(K_1^L+\phi_u)|I\rangle$, where $|I\rangle$ is the identity string field (and we remark that in our applications $\phi_u(\tilde z)$ is always inserted in the left part of the string). Therefore the inverse of $K+\phi_u$ can also be obtained
via the inverse of the operator $ K_1^L+\phi_u$.

The operator ${\cal K}_u\equiv  K_1^L+\phi_u$ is a self-adjoint operator. Therefore its spectrum lies on the real axis. To know more about it we would need a spectral analysis of ${\cal K}_u$, similar to what has been done for the operator $K_1^L$ in \cite{RSZ,BMST1,BMST2,BMT3}. The spectrum of the latter is the entire real axis. The spectrum of ${\cal K}_u $ is of course expected to be different, but we know on a general ground that it lies on the real axis. We can therefore define the resolvent
of ${\cal K}_u$, $R(\kappa, {\cal K}_u)$, which is by definition the inverse of $\kappa- {\cal K}_u$. The resolvent is well defined (at least) for any non-real $\kappa$.
We do not know what type of eigenvalue the $\kappa=0$ one is: discrete, continuous or residual. However, since $R(\kappa, {\cal K}_u) (\kappa- {\cal K}_u)=1$ is true for any $\kappa$
outside the real axis, we can hold it valid also in the limit $\kappa\to 0$ by continuity\footnote{
To the risk of pedantry, let us consider a simple model of the situation. In the complex $z$ plane 
the expression $\frac 1z$ is of course singular at $z=0$, but the expression $z \cdot \frac 1z$  
is 1 on the whole complex plane (not only for $z\neq 0$), in virtue of continuity. Naturally if one evaluates $\frac 1z$ first at $z=0$ and then multiplies it by 0, one ends up with an indefinite expression, but this is the wrong way to proceed.}.

On a general ground we can therefore conclude that eq.(\ref{IDD}) must be true. Then, how do we explain the discrepancy between \eqref{LHS} and \eqref{wrong}? These two equations are affected by an UV singularity and need a subtraction. Without this subtraction they are meaningless.
We have seen in this paper (more examples can be found in \cite{BGT2}) that there is no canonical subtraction scheme for such singularities, and a physical meaning can be assigned only to quantities that are subtraction-independent. The difference between the RHS of (\ref{LHS}) and (\ref{wrong}) is of course due to the use of a Schwinger representation utilized to derive the second, similar to what we have seen in the calculation of the energy. The Schwinger representation allows us to transplant in (\ref{wrong}) a path integral result (the partition function $g(t)$ calculated by Witten). In view of this it would actually be surprising to find 
the same result in the RHS of (\ref{LHS}) and (\ref{wrong}). Nevertheless (\ref{LHS}) and (\ref{wrong}) 
must represent the same thing. Since $g(\infty)$ is finite, we can figure out a subtraction scheme 
for  (\ref{LHS}) and another for (\ref{wrong}) so that the subtracted quantities
coincide (if we subtract an infinity we can also subtract an infinity plus something finite). 
If we need to represent the identity as in the LHS of (\ref{LHS}) we will use one subtraction, if instead we need
to represent it as in the LHS of (\ref{wrong}) we will use another.
Thus, in principle we can eliminate any contradiction between  (\ref{LHS}) and (\ref{wrong}). However the next question is whether the use of two subtractions schemes is compatible with the validity of the equation of motion. We have not been able to prove or disprove this in a convincing way, therefore we prefer to take a conservative point of view. 

\subsection{A conservative viewpoint}

The equation of motion for $\psi_u$ is true on the basis of the previous general argument. As a consequence
we have to conclude that the Schwinger representation may not be reliable when applied to the equation of motion
and we will avoid it. But if the Schwinger representation is not essential (and perhaps inadequate) when applied to the equation of motion, it is essential in the energy calculation. Thus, how can we trust the result we have obtained in section 8 and 9? We will argue in this last part of the appendix that when applied to the energy calculation
the Schwinger representation gives a consistent result, provided we drop the offending term proportional to $\Omega_u^\infty$ in (\ref{wrong}).  

To start with let us remark that our end result (\ref{lumpenergy}), obtained via a coherent procedure, is the expected one. This cannot be explained unless the 
calculation of the energy by means of the Schwinger representation, barring miraculous cancellations, is the correct one.
Secondly, as we have remarked above, the ambiguity (if we can call it this way) of the Schwinger representation, that is (\ref{wrong}) as compared to (\ref{LHS}), involves the UV subtraction. But we have shown above that the energy calculation is independent of the subtraction scheme used. Therefore we conclude that our result must not be affected by the ambiguity. 

The above arguments are, however, indirect. A more direct one is the following.
From (\ref{1-Omega}) we can write formally
\be
\frac 1{K+\phi_u}= \int_0^\infty dt\, e^{-t(K+\phi_u)}+ \frac 1{K+\phi_u} \Omega_u^\infty
\label{Schwrong}
\ee
On the other hand from section 7 we have
\be
\frac 1{K+\phi_u+\epsilon}= \int_0^\infty dt\, e^{-t(K+\phi_u+\e)}\label{Schright}
\ee
Our solution is 
\be 
\psi_{u}=c\phi_u-\frac1{K+\phi_u}(\phi_u-\delta\phi_u) Bc\del c\label{psiphiu} 
\ee
In computing the energy of $\psi_u$ (section 2) we have utilized only the first part
of the RHS of (\ref{Schwrong}) and disregarded the second part (the already mentioned Schwinger ambiguity). This is intuitively correct: we have just dropped the term that may violate the equation of motion. It is nevertheless a move not automatically inscribed in the formalism, therefore, as logical as it may seem, it needs some justification. To justify it we will show that the energy obtained
in this way is the limit of a regularized version of the energy functional where the
use of the Schwinger representation is justified beyond ambiguity.  

To this end let us consider the following deformation of (\ref{psiphiu}):
\be 
\psi_{u,\e}=c\phi_u-\frac1{K+\phi_u+\e}(\phi_u-\delta\phi_u) Bc\del c\label{psiphiue} 
\ee
This is not a solution to the equation of motion, but nothing prevents us from using
it as an auxiliary string field. Let us do the following exercise: let us replace
everywhere in (\ref{TVlumpenergy}) $\psi_u$ with $\psi_{u,\e}$. In all the 
correlators in (\ref{TVlumpenergy}) we have to use (\ref{Schright}). This induces some changes in the formulas of section 2 and 9.1, but is easy to see that in the limit
$\e\to 0$ the result coincides piece by piece. In the discussion in section 8 it has already been remarked that $\langle \psi_u,\psi_u \psi_u\rangle$ is the same
as $\langle \psi_{u,\e},\psi_{u,\e} \psi_{u,\e}\rangle$ in the limit $\e\to 0$ (see in particular the paragraph around eq.(\ref{whatremains})). The same relation holds between
$\langle \psi_{\e},\psi_{u,\e} \psi_{u,\e}\rangle$ and $\langle \psi_{\e},\psi_{u} \psi_{u}\rangle$  and between $\langle \psi_{\e},\psi_{\e} \psi_{u,\e}\rangle$ and $\langle \psi_{\e},\psi_{\e} \psi_{u}\rangle$. The relevant correlators with $\psi_{u,\e}$ instead of $\psi_u$ are obtained from eqs.(\ref{Fs2},\ref{Fs3}) by suppressing the factors $e^{-\eta sy}$ and $e^{-\eta sx}$ in the latter. In the limit
$\e\to 0$ the equalities are thus established.

In conclusion, Eq.(\ref{TVlumpenergy}) is the same whether we compute it by using
for $\psi_u$ the Schwinger representation (\ref{Schwrong}) and dropping the $\Omega_u^\infty$ piece (as we have done throughout this paper), or we replace
$\psi_u$ with $\psi_{u,\e}$ and use the regularized Schwinger representation
(\ref{Schright}) in the limit $\e\to 0$. In other words we have achieved a representation of the energy functional where the use of Schwinger representation (\ref{Schright}) does not suffer from any ambiguity and its use is absolutely legitimate. 
The outcome of this discussion is that the offending term proportional to $\Omega_u^\infty$ in (\ref{wrong},\ref{Schwrong}) must not be taken into consideration when computing the energy.

In conclusion, in this Appendix, which is actually a prolongation of \cite{BMT}, we have seen that the Schwinger representation may not provide a faithful representation of $\frac 1{K+\phi_u}(K+\phi_u)=1$ and may affect the
proof of the equation of motion (although we have not been able either to prove or disprove this point). 
We have however argued that the calculation of the energy by means of the Schwinger
representation we have used throughout this paper is perfectly consistent.


\begin{thebibliography}{99}

\bibitem{BMT}
  L.~Bonora, C.~Maccaferri and D.~D.~Tolla,
 {\it Relevant Deformations in Open String Field Theory: a Simple Solution for
  Lumps,}
  arXiv:1009.4158 [hep-th].

\bibitem{Witten:1985cc}
  E.~Witten,
  {\it Noncommutative Geometry And String Field Theory,}
  Nucl.\ Phys.\  B {\bf 268} (1986) 253.
 
\bibitem{Sen:1999mh}
  A.~Sen,
  ``Descent relations among bosonic D-branes,''
  Int.\ J.\ Mod.\ Phys.\  A {\bf 14}, 4061 (1999)
  [arXiv:hep-th/9902105].

\bibitem{Sen:1999xm}
  A.~Sen,
  ``Universality of the tachyon potential,''
  JHEP {\bf 9912}, 027 (1999)
  [arXiv:hep-th/9911116].

\bibitem{Schnabl05}
  M.~Schnabl,
  {\it Analytic solution for tachyon condensation in open string field theory,}
  Adv.\ Theor.\ Math.\ Phys.\  {\bf 10} (2006) 433
  [arXiv:hep-th/0511286].

\bibitem{Okawa1}
  Y.~Okawa,
 {\it Comments on Schnabl's analytic solution for tachyon condensation in
  Witten's open string field theory,}
  JHEP {\bf 0604} (2006) 055
  [arXiv:hep-th/0603159].


\bibitem{ErlerSchnabl}
  T.~Erler and M.~Schnabl,
  {\it A Simple Analytic Solution for Tachyon Condensation,}
  arXiv:0906.0979 [hep-th].

\bibitem{RZ06}
  L.~Rastelli and B.~Zwiebach,
  {\it Solving open string field theory with special projectors,}
arXiv:hep-th/0606131.

\bibitem{ORZ}
  Y.~Okawa, L.~Rastelli and B.~Zwiebach,
 {\it Analytic solutions for tachyon condensation with general projectors,}
  arXiv:hep-th/0611110.

\bibitem{Fuchs0}
  E.~Fuchs and M.~Kroyter,
 {\it On the validity of the solution of string field theory,}
  JHEP {\bf 0605} (2006) 006
  [arXiv:hep-th/0603195].

\bibitem{Erler:2006hw}
  T.~Erler,
  {\it Split string formalism and the closed string vacuum},
  JHEP {\bf 0705}, 083 (2007)
  [arXiv:hep-th/0611200].

\bibitem{Erler:2006ww}
  T.~Erler,
  {\it Split string formalism and the closed string vacuum. II},
  JHEP {\bf 0705}, 084 (2007)
  [arXiv:hep-th/0612050].

\bibitem{Erler:2007xt}
  T.~Erler,
  {\it Tachyon Vacuum in Cubic Superstring Field Theory,}
  JHEP {\bf 0801} (2008) 013
  [arXiv:0707.4591 [hep-th]].


\bibitem{Arroyo:2010fq}
  E.~A.~Arroyo,
  {\it Generating Erler-Schnabl-type Solution for Tachyon Vacuum in Cubic
  Superstring Field Theory},
  arXiv:1004.3030 [hep-th].


\bibitem{Zeze:2010jv}
  S.~Zeze,
  {\it Tachyon potential in KBc subalgebra},
  arXiv:1004.4351 [hep-th].

\bibitem{Zeze:2010sr}
  S.~Zeze,
  {\it Regularization of identity based solution in string field theory},
  arXiv:1008.1104 [hep-th].

\bibitem{Arroyo:2010sy}
  E.~A.~Arroyo,
  {\it Comments on regularization of identity based solutions in string field
  theory,}
  arXiv:1009.0198 [hep-th].

\bibitem{Murata}
  M.~Murata and M.~Schnabl,
  {\it On Multibrane Solutions in Open String Field Theory,}
  Prog.\ Theor.\ Phys.\ Suppl.\  {\bf 188}, 50 (2011)
  [arXiv:1103.1382 [hep-th]].


  \bibitem{KORZ}
  M.~Kiermaier, Y.~Okawa, L.~Rastelli and B.~Zwiebach,
 {\it Analytic solutions for marginal deformations in open string field theory,}
  arXiv:hep-th/0701249.

  \bibitem{Schnabl:2007az}
  M.~Schnabl,
 {\it Comments on marginal deformations in open string field theory,}
  arXiv:hep-th/0701248.

\bibitem{Kluson:2003xu}
  J.~Kluson,
{\it Exact solutions in SFT and marginal deformation in BCFT,}
  JHEP {\bf 0312}, 050 (2003).
  [hep-th/0303199].

\bibitem{Kiermaier:2007vu}
  M.~Kiermaier and Y.~Okawa,
  {\it Exact marginality in open string field theory: a general framework,}
  arXiv:0707.4472 [hep-th].


\bibitem{Fuchs3}
  E.~Fuchs, M.~Kroyter and R.~Potting,
  {\it Marginal deformations in string field theory,}
   arXiv:0704.2222 [hep-th].


\bibitem{Lee:2007ns}
  B.~H.~Lee, C.~Park and D.~D.~Tolla,
  {\it  Marginal Deformations as Lower Dimensional D-brane Solutions in Open String
  Field theory,}
  arXiv:0710.1342 [hep-th].


\bibitem{Kwon:2008ap}
  O.~K.~Kwon,
  {\it Marginally Deformed Rolling Tachyon around the Tachyon Vacuum in Open
  String Field Theory},
  Nucl.\ Phys.\  B {\bf 804}, 1 (2008)
  [arXiv:0801.0573 [hep-th]].

  \bibitem{Okawa2}
  Y.~Okawa,
{\it Analytic solutions for marginal deformations in open
superstring field
  theory,}
  arXiv:0704.0936 [hep-th].

\bibitem{Okawa3}
  Y.~Okawa,
{\it Real analytic solutions for marginal deformations in open
superstring field theory,}
  arXiv:0704.3612 [hep-th].





\bibitem{Kiermaier:2007ki}
  M.~Kiermaier and Y.~Okawa,
  {\it General marginal deformations in open superstring field theory,}
  arXiv:0708.3394 [hep-th].



\bibitem{Erler:2007rh}
  T.~Erler,
  {\it Marginal Solutions for the Superstring,}
  JHEP {\bf 0707} (2007) 050
  [arXiv:0704.0930 [hep-th]].


  \bibitem{Fuchs4}
E.~Fuchs and M.~Kroyter, {\it Analytical Solutions of Open String
Field Theory,}
  arXiv:0807.4722 [hep-th].

\bibitem{Schnabl:2010tb}
  M.~Schnabl,
  {\it Algebraic solutions in Open String Field Theory - a lightning review},
  arXiv:1004.4858 [hep-th].
  \bibitem{lumps}
  N.~Moeller, A.~Sen and B.~Zwiebach,
  {\it D-branes as tachyon lumps in string field theory,}
  JHEP {\bf 0008} (2000) 039
  [arXiv:hep-th/0005036].

\bibitem{Witten}
  E.~Witten,
 {\it Some computations in background independent off-shell string theory,}
  Phys.\ Rev.\  D {\bf 47}, 3405 (1993)
  [arXiv:hep-th/9210065].

\bibitem{Kutasov}
  D.~Kutasov, M.~Marino and G.~W.~Moore,
{\it Some exact results on tachyon condensation in string field
theory,}
  JHEP {\bf 0010}, 045 (2000)
  [arXiv:hep-th/0009148].

\bibitem{Ellwood} I.~Ellwood,
  {\it Singular gauge transformations in string field theory,}
  JHEP {\bf 0905}, 037 (2009)
  [arXiv:0903.0390 [hep-th]].

\bibitem{EllwoodSchnabl}
  I.~Ellwood and M.~Schnabl,
 {\it Proof of vanishing cohomology at the tachyon vacuum,}
  JHEP {\bf 0702} (2007) 096
  [arXiv:hep-th/0606142].

\bibitem{RSZ}
  L.~Rastelli, A.~Sen and B.~Zwiebach,
 {\it Star algebra spectroscopy,}
  JHEP {\bf 0203}, 029 (2002)
  [arXiv:hep-th/0111281].


\bibitem{BMST1}
  L.~Bonora, C.~Maccaferri, R.~J.~Scherer Santos and D.~D.~Tolla,
  {\it Ghost story. I. Wedge states in the oscillator formalism,}
  JHEP {\bf 0709}, 061 (2007)
  [arXiv:0706.1025 [hep-th]].

\bibitem{BMST2}
  L.~Bonora, C.~Maccaferri, R.~J.~Scherer Santos and D.~D.~Tolla,
 {\it Ghost story. II. The midpoint ghost vertex,}
  JHEP {\bf 0911}, 075 (2009)
  [arXiv:0908.0055 [hep-th]].

\bibitem{BMT3}
  L.~Bonora, C.~Maccaferri and D.~D.~Tolla,
{\it Ghost story. III. Back to ghost number zero,}
  JHEP {\bf 0911}, 086 (2009)
  [arXiv:0908.0056 [hep-th]].

\bibitem{DS} Dunford and Schwartz, {\it Linear Operators}, vol. I,II.

\bibitem{Polchinski} J.Polchinski, {\it String Theory. Volume I}, Cambridge University Press, Cambridge 1998.

\bibitem{Okawa02}
  Y.~Okawa,
  {\it Open string states and D-brane tension from vacuum string field theory,}
  JHEP {\bf 0207} (2002) 003
  [arXiv:hep-th/0204012].

\bibitem{EM}
  T.~Erler and C.~Maccaferri,
 {\it Comments on Lumps from RG flows,}
  arXiv:1105.6057 [hep-th].

 
\bibitem{BGT2}
  L.~Bonora, S.~Giaccari and D.~D.~Tolla,
{\it Analytic solutions for Dp branes in SFT,} JHEP12(2011)033.
  arXiv:1106.3914 [hep-th].

\bibitem{BGT3}
L.~Bonora, S.~Giaccari and D.~D.~Tolla, {\it Lump solutions in SFT. Complements}, hep-th/1109.4336, {\it expanded version}
  
\end{thebibliography}
\end{document}